\documentclass[fleqn,usenatbib,letters]{mnras}
\usepackage[T1]{fontenc}
\usepackage{graphicx}
\usepackage{amsmath}
\usepackage{amssymb}
\usepackage{xcolor}
\usepackage{hyperref}
\graphicspath{{figures/}}

\newcommand{\feh}{[{\rm Fe/H}]}
\newcommand{\kms}{\,km\,s$^{-1}$}

\newcommand{\Msun}{\,M$_\odot$}

\newcommand{\ltFehA}{-1.5}
\newcommand{\ltFehB}{-1.1}
\newcommand{\ltFehC}{-0.8}
\newcommand{\ltFehD}{-0.5}
\newcommand{\ltRingIn}{2}
\newcommand{\ltRingOut}{9}
\newcommand{\ltRrlFitIn}{4}

\newcommand{\ltNBoss}{49\,712}
\newcommand{\ltNApogee}{6\,719}

\newcommand{\ltProfIn}{3}
\newcommand{\ltProfOut}{7}

\newcommand{\ltRSig}{4}
\newcommand{\ltApoFeh}{-1.1}
\newcommand{\ltNTng}{681}
\newcommand{\ltNAu}{5}
\newcommand{\ltNFire}{3}
\newcommand{\ltMstarLo}{1.5}
\newcommand{\ltMstarHi}{5}
\newcommand{\ltMtotLo}{5\times10^{10}}
\newcommand{\ltMtotHi}{5\times10^{11}}

\newcommand{\ltDist}{50.1}
\newcommand{\ltFirstFrac}{2}

\newcommand{\ltCentreOffset}{0.5}
\newcommand{\ltKinVrotMP}{57}
\newcommand{\ltKinVrotMR}{78}
\newcommand{\ltKinSigMP}{32}
\newcommand{\ltKinSigMR}{16}
\newcommand{\ltKinRatioMP}{1.8}
\newcommand{\ltKinRatioMR}{4.9}
\newcommand{\ltKinRatioMRApo}{7.7}

\newcommand{\ltKinRatioRMP}{1.5}

\newcommand{\ltKinRrlRatioR}{1.7}

\newcommand{\ltTngRatioMP}{0.3}

\newcommand{\ltTngRatioMPMax}{1.0}

\newcommand{\ltAuRatioMRMax}{4.3}
\newcommand{\ltFireRatioMRMax}{3.4}

\newcommand{\ltTngPctC}{24}
\newcommand{\ltTngPctD}{49}
\newcommand{\ltTngPctE}{20}
\newcommand{\ltEarlyMass}{5\times10^{7}}
\newcommand{\ltEarlyMassLo}{4}
\newcommand{\ltEarlyMassHi}{7}
\newcommand{\ltEarlyExp}{7}
\newcommand{\ltEarlyRatio}{40}
\newcommand{\ltEarlyRatioMW}{10}
\newcommand{\ltMPFehMed}{-1.8}
\newcommand{\ltMPFehLow}{-2}
\newcommand{\ltMPPctLow}{25}
\newcommand{\ltLowFeh}{-2}
\newcommand{\ltLowN}{562}
\newcommand{\ltLowMass}{10^{7}}
\newcommand{\ltLowRatioLo}{1.5}
\newcommand{\ltLowRatioHi}{1.7}
\newcommand{\ltLowPurity}{40}
\newcommand{\ltLowPurityMin}{20}
\newcommand{\ltLowInterMax}{80}
\newcommand{\ltLowInterLo}{-1.9}
\newcommand{\ltLowInterHi}{-1.7}

\newcommand{\ltHMP}{1.9}
\newcommand{\ltReMP}{3.2}
\newcommand{\ltHMR}{0.9}
\newcommand{\ltReMR}{1.5}
\newcommand{\ltHRrl}{1.7}
\newcommand{\ltReRrl}{2.9}
\newcommand{\ltHFitLo}{2}
\newcommand{\ltHFitHi}{8}

\newcommand{\ltTngColdVrot}{3}
\newcommand{\ltOldAgeLo}{10}
\newcommand{\ltOldAgeHi}{12}
\newcommand{\ltTngRbMed}{1.7}

\newcommand{\ltTngRtMed}{0.3}

\newcommand{\ltTngVbMed}{35}
\newcommand{\ltTngVtMed}{11}
\newcommand{\ltTngSbMed}{20}
\newcommand{\ltTngStMed}{41}
\newcommand{\ltAuRbMed}{1.4}
\newcommand{\ltAuRtMed}{0.2}
\newcommand{\ltKinRatioB}{1.9}
\newcommand{\ltNTngBirth}{681}

\newcommand{\ltRToday}{10}
\newcommand{\ltCohort}{0.1}

\newcommand{\ltRBirth}{3}
\newcommand{\ltRMinBirth}{0.5}

\newcommand{\ltNFull}{20}

\newcommand{\ltZSnapA}{4}
\newcommand{\ltZSnapB}{1.5}
\newcommand{\ltWindow}{25}

\newcommand{\ltDtMax}{0.5}
\newcommand{\ltNCohorts}{92}
\newcommand{\ltNCohortGal}{16}

\newcommand{\ltROrb}{2}

\newcommand{\ltTOrb}{0.2}
\newcommand{\ltTAfterA}{0.2}
\newcommand{\ltTAfterB}{0.35}

\newcommand{\ltSetRZero}{2}
\newcommand{\ltSetRIn}{3}
\newcommand{\ltNSample}{60}

\newcommand{\ltScrStart}{1.9}
\newcommand{\ltScrSecond}{1.0}
\newcommand{\ltScrEnd}{0.6}
\newcommand{\ltScrTSecond}{0.15}
\newcommand{\ltScrTEnd}{0.4}

\newcommand{\ltFrozenEnd}{0.7}
\newcommand{\ltDmEnd}{0.5}
\newcommand{\ltAxiEnd}{1.9}
\newcommand{\ltGasRatio}{1.8}
\newcommand{\ltEpsLow}{0.005}
\newcommand{\ltEpsHigh}{0.02}
\newcommand{\ltVsEpsLow}{1.7}
\newcommand{\ltVsEpsHigh}{0.05}
\newcommand{\ltEpsSpearman}{-0.7}
\newcommand{\ltNCohortsB}{92}
\newcommand{\ltSettleLo}{2}
\newcommand{\ltSettleHi}{4}
\newcommand{\ltLbEarly}{12.5}
\newcommand{\ltLbLate}{7.5}
\newcommand{\ltPsiEarly}{40}
\newcommand{\ltPsiLate}{10}
\newcommand{\ltEpsEarly}{0.02}
\newcommand{\ltEpsLate}{0.006}
\newcommand{\ltCaEarly}{0.75}
\newcommand{\ltCaLate}{0.84}

\newcommand{\ltSigmaZero}{147}
\newcommand{\ltWSidm}{120}

\newcommand{\ltREps}{2}

\newcommand{\ltSiCdmNaLo}{0.008}
\newcommand{\ltSiCdmNaHi}{0.02}
\newcommand{\ltSiEpsFirst}{0.004}

\newcommand{\ltSiEpsLate}{0.0004}
\newcommand{\ltSiRatioLate}{25}
\newcommand{\ltSiCaCdmLo}{0.4}
\newcommand{\ltSiCaCdmHi}{0.5}

\newcommand{\ltSiCaSidmMed}{0.9}
\newcommand{\ltSiZFirst}{4}
\newcommand{\ltRadPartRatio}{0.2}

\newcommand{\ltRadBornOuter}{20}

\newcommand{\ltFireRbMed}{1.0}

\title[The puzzle of the LMC's ancient disc]{The puzzle of the ancient stellar disc of the Large Magellanic Cloud}
\author[V. Belokurov]{V. Belokurov$^{1}$\thanks{E-mail: vasily@ast.cam.ac.uk}\\
$^{1}$Institute of Astronomy, University of Cambridge, Madingley Road, Cambridge CB3 0HA, UK}
\date{Accepted XXX. Received YYY; in original form ZZZ}
\pubyear{2026}

\begin{document}
\label{firstpage}
\pagerange{\pageref{firstpage}--\pageref{lastpage}}
\maketitle

\begin{abstract}
We measure the rotational support of the Large Magellanic Cloud's
(LMC) stars against metallicity, from the three-dimensional velocities
of APOGEE and SDSS-V red giants with \textit{Gaia} proper motions. The LMC is a rotating disc at every
metallicity, and its RR Lyrae are consistent with this. Its giants with
$\feh<\ltFehA$ have $v_{\rm rot}/\sigma_z\approx\ltKinRatioMP$, and even
those with $\feh<\ltLowFeh$ have $\ltLowRatioLo<v_{\rm rot}/\sigma_z<\ltLowRatioHi$.
The LMC had this disc after forming at most $\sim\ltEarlyMass$\Msun\ of stars, some
\ltEarlyRatio\ times below the mass at which stable
discs emerge in most simulations. No LMC analogue in TNG50,
Auriga or FIRE-2 matches the rotational support of its oldest stars. In TNG50
and Auriga these stars are born rotating but halve their rotation
within one orbital period and lose most of it within two: torques from
the aspherical, misaligned dark-matter halo scramble their orbits. Discs survive only once
the halo's minor axis aligns with the disc's spin and it becomes
rounder, after the LMC's oldest stars formed. In a self-interacting
dark-matter resimulation of an LMC-mass halo, the inner halo is
already nearly round at $z\approx\ltSiZFirst$, which could let early discs
survive.
\end{abstract}


\section{Introduction}

The RR Lyrae of the Large Magellanic Cloud (LMC) follow the spatial
distribution of its disc giants closely: the same extent, the same
shape, and even the same sharp truncation on the side facing the Small
Magellanic Cloud \citep[SMC;][fig.~1]{Grady2021}. Yet they are old and
metal-poor \citep{Alcock1996,Olsen1998,Gratton2004,Borissova2006,Skowron2016}.
If they belong to the disc, the LMC has had a disc since its oldest
stars formed.

That would test theories of disc emergence. In most simulations, discs
appear only above a stellar mass of $\sim10^9$\Msun, whatever the
epoch; below it, the stars are supported by random motions
\citep{ElBadry2018,Clauwens2018,Dekel2020,Semenov2025a}. Two kinds of
explanation are proposed. Either the gas cannot keep its spin, because
mergers and changing inflows keep turning it, so that stars born at
different times cancel each other's rotation
\citep{McCarthy2012,Sales2012,Dekel2020,Dillamore2024,Semenov2024b,Zeng2024};
or a disc can settle only once the galaxy has grown a hot gaseous halo
or a centrally concentrated mass profile
\citep{Stern2021,Hafen2022,Hopkins2023,Semenov2026}.
Curiously, in the Marvelous Massive Dwarfs simulations, discs can
form in dwarfs down
to $M_\star\approx10^{7.7}$\Msun\ when gas-rich satellites on prograde
orbits spin up the gas, and grow from the stars formed after these
mergers \citep{Geda2026}.

\begin{figure*}
\includegraphics[width=\textwidth]{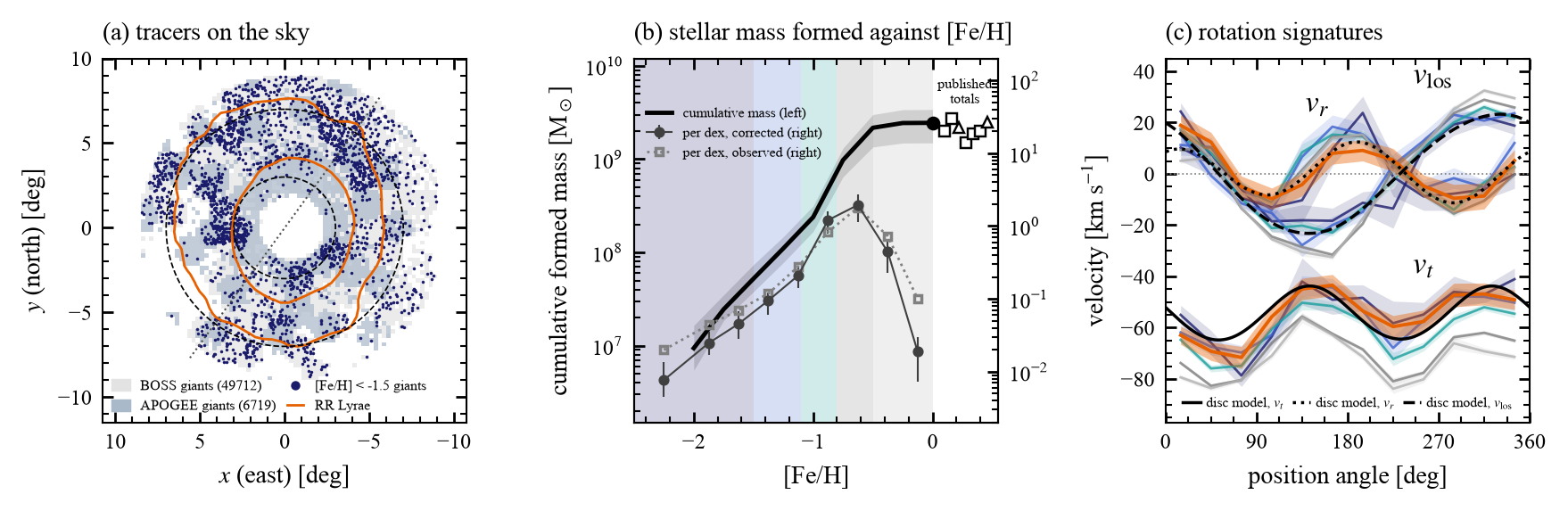}
\caption{The LMC's stellar tracers, the mass they formed and their
  rotation. (a) The tracers on the sky about the fitted centre: pixels
  with BOSS (grey) and APOGEE (slate) giants, giants with
  $\feh<\ltFehA$ (dark blue dots), RR Lyrae density (orange contours);
  dashed circles: \ltProfIn$^\circ$ and \ltProfOut$^\circ$ from the
  fitted centre; dotted: the line of nodes. (b) Stellar mass formed
  against \feh: cumulative (left axis; filled circle: the total;
  triangles and squares: published formed and present masses) and
  fraction per dex (right axis; circles: corrected, open squares:
  observed giants); shading: the five metallicity bins. (c) Median
  $v_t$, $v_r$ (all populations) and $v_{\rm los}$ (giants) against
  position angle \ltProfIn--\ltProfOut$^\circ$ from the fitted centre;
  colours as the shading in (b), RR Lyrae orange; black: the reference
  disc ($v_t$ solid, $v_r$ dotted, $v_{\rm los}$ dashed). Every
  population shows the velocity pattern of a rotating disc.
}
\label{fig:data}
\end{figure*}

Kinematic studies of galaxies at high redshift are at present limited
to their gas.
Ionised and cold gas give different rotational support
\citep{Simons2016,Rizzo2020,Girard2021,deGraaff2024,Rizzo2024,Rowland2024,Kohandel2024,Danhaive2025}.
A rotating gas disc also need not leave a coherently rotating stellar
disc: stars form with less rotational support than their gas, and if
the gas disc changes its orientation, stars born at different times do
not share one rotation \citep{ElBadry2018,Zeng2024,Kohandel2025}.
Archaeological studies of nearby galaxies measure the stars
themselves. The Milky Way spun up its disc between $\feh\approx-1.3$
and $-0.9$, after forming $\sim5\times10^8$\Msun\ of stars
\citep[][see also \citealt{Conroy2022,Xiang2022,Rix2022}]{Belokurov2022,Belokurov2023,Kurbatov2024},
but most dwarf galaxies with
$10^{3.5}<M_\star<10^{8}$\Msun\ are supported by dispersion today, in isolation
as well as near massive hosts \citep{Wheeler2017}, and even dwarfs in voids
with $M_\star=10^7$--$10^9$\Msun\ have a ratio of rotation to
dispersion $v/\sigma\lesssim2$ \citep{delosReyes2023}, although such
measurements, dominated by old stars and by spatially binned spectra,
may underestimate $v/\sigma$ \citep{Ruan2026}.

The LMC lies between the dispersion-supported dwarf spheroidals and the
Milky Way. It is massive enough for its oldest stars to be studied in
large numbers, and its record is cleaner than the Milky Way's, whose
early disc stars were thrown onto halo-like orbits by its last significant
merger \citep{Bonaca2017,DiMatteo2019,Gallart2019,Belokurov2020}: dwarf galaxies form almost all their
stars in situ \citep[][but see \citealt{Mucciarelli2021}]{Fitts2018}.

Earlier work suggests that the LMC's old populations belong to its
disc. Its old clusters form a flattened, rotating system with no
kinematic halo \citep{Freeman1983,Schommer1992}, carbon stars trace a
thick rotating disc \citep{vanderMarel2002}, the surface brightness and
main-sequence star counts stay exponential to large radii
\citep{Gallart2004,Saha2010}, and the most metal-poor red giants lie in
a thicker disc rather than a hot halo \citep{Carrera2011}. Halo
candidates exist too. The density of the RR Lyrae admits a halo
\citep{Kinman1991}, and RR Lyrae in the inner LMC have a high
line-of-sight velocity dispersion \citep{Minniti2003,Pejcha2009}; an
extended, lumpy envelope surrounds the LMC
\citep{Majewski2009,Belokurov2016,Belokurov2017,Navarrete2019}, but it
holds only $\sim$0.4 per cent of its stellar mass \citep{Nidever2019}
and may be its outer disc, disturbed by the SMC and the Milky Way
\citep{Mackey2016,Mackey2018,Cullinane2022a,Cullinane2022b,Navarrete2023}.

However, RR Lyrae are not the best tracers for the question: for most
of them only proper motions are available, which do not measure the
vertical velocity dispersion. We therefore use red giants with spectra
from APOGEE \citep{Majewski2017,Abdurrouf2022} and SDSS-V
\citep{Kollmeier2026,Nidever2026} and proper motions from
\textit{Gaia} DR3 \citep{GaiaCollaboration2023} to measure all three
components of their velocities, and the rotational support of the
LMC's stars as a function of metallicity, and hence of the stellar
mass the LMC had formed when they were born. The RR Lyrae serve as a
check, and we compare the result with LMC analogues in the TNG50,
Auriga and FIRE-2 simulations \citep{Pillepich2019,Grand2024,Wetzel2023}.

\section{Rotational support across metallicity}\label{sec:rotation}

We use the red giants of two surveys, APOGEE DR17
\citep{Majewski2017,Abdurrouf2022} and SDSS-V DR20 BOSS
\citep{Kollmeier2026,Sizemore2024,Nidever2026}, with \textit{Gaia} DR3
proper motions \citep{GaiaCollaboration2023} and line-of-sight
velocities, $v_{\rm los}$, and, as a check, the fundamental-mode (RRab)
RR Lyrae of OGLE-IV and \textit{Gaia} DR3
\citep{Udalski2015,Soszynski2016,Soszynski2019,Clementini2023}, with
proper motions only. The giants are split
into five \feh\ bins ($<\ltFehA$, $\ltFehA$ to $\ltFehB$, $\ltFehB$ to
$\ltFehC$, $\ltFehC$ to $\ltFehD$, $>\ltFehD$), and all stars are kept
at $\ltRingIn^\circ<\rho<\ltRingOut^\circ$, where $\rho$ is the angular
distance from the centre of \citet{vanderMarel2014}; velocities and
maps are measured about the centre fitted by our disc model (the
fitted centre), \ltCentreOffset$^\circ$ east of it.

Fig.~\ref{fig:data}(a) shows the sky coverage of each tracer. APOGEE (\ltNApogee\ giants) observed separate fields, BOSS
(\ltNBoss) fills most of the disc, and the RR Lyrae cover it almost
without interruption. Fig.~\ref{fig:data}(b) is the stellar mass formed
per \feh, from the counts of giants corrected for selection, coverage,
the Milky Way foreground, the scatter of BOSS \feh, and the number of
giants per unit formed mass (PARSEC isochrones,
\citealt{Bressan2012,Marigo2017}; Kroupa IMF, \citealt{Kroupa2001};
ages from the age--metallicity relation of \citealt{Povick2024}). For
each metallicity we thus know how much stellar mass the LMC had formed
before stars of that metallicity were born, and, through the
age--metallicity relation, roughly when.

\begin{figure*}
\includegraphics[width=\textwidth]{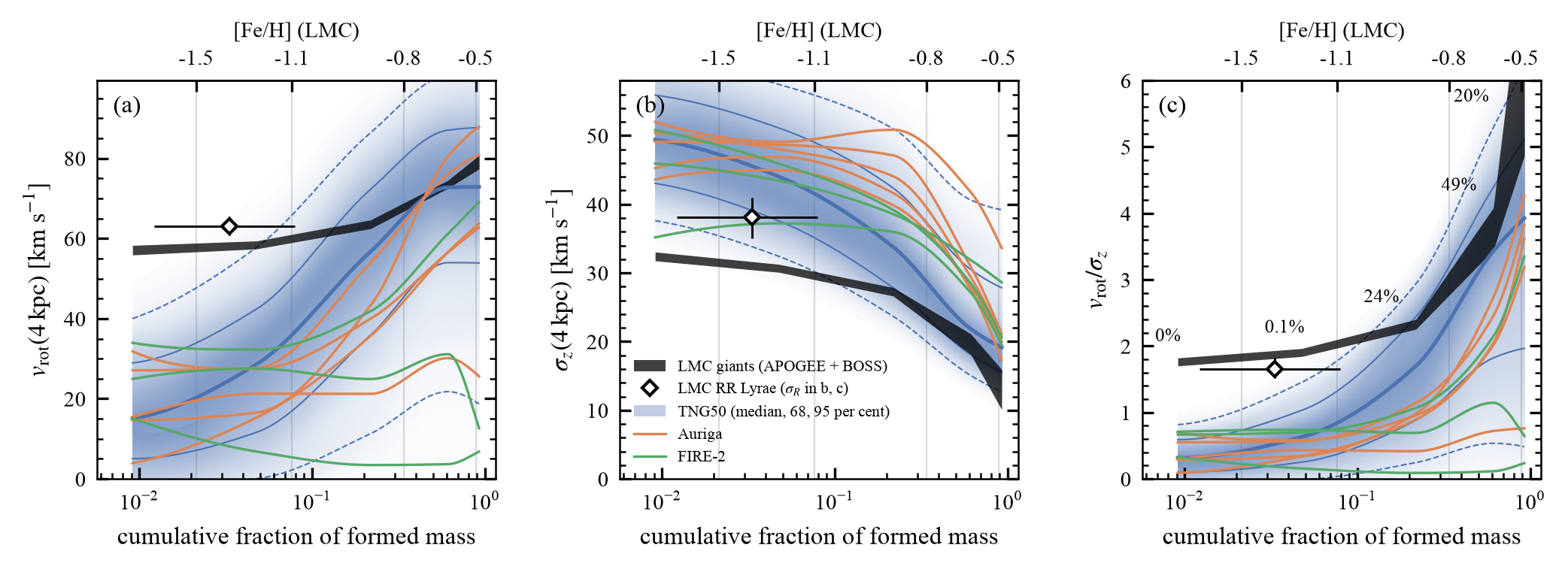}
\caption{The rotational support of the LMC's stars against that of
  simulated analogues observed and fitted in the same way, at
  $R=\ltRSig$\,kpc: (a) $v_{\rm rot}$, (b) $\sigma_z$, (c)
  $v_{\rm rot}/\sigma_z$, against the cumulative fraction of the formed
  mass below each metallicity bin; top axis: the bin edges in LMC
  \feh. Black bands: the LMC giants (16th--84th percentiles; above
  $\feh=\ltApoFeh$ extended to APOGEE alone, hence the widening). Blue:
  \ltNTng\ TNG50 analogues (median, thick; 68 and 95 per cent ranges,
  thin and dashed); orange: \ltNAu\ Auriga, green: \ltNFire\ FIRE-2
  analogues. Numbers in (c): per cent of TNG50 analogues reaching the
  LMC band. Diamonds: the RR Lyrae, with $\sigma_R$ in (b) and (c).
  Almost no analogue is as rotation-supported as the LMC at
  $\feh<\ltFehB$; only TNG50 matches its metal-rich stars.
}
\label{fig:sims_kin}
\end{figure*}

Fig.~\ref{fig:data}(c) shows the internal velocities, with only the
LMC's bulk motion removed: from each star's proper motion and
$v_{\rm los}$ we subtract those that the LMC's centre-of-mass motion
would give at its position, in the frame of a joint fit of all
metallicity bins with the disc model below. The proper motions, converted to velocities
at each star's distance in the fitted disc plane, are split into
components along ($v_r$) and perpendicular to ($v_t$, negative
clockwise on the sky) the direction away from the fitted centre. In
every metallicity bin of the giants $v_{\rm los}$ makes one wave per
turn in position angle, and in every bin and for the RR Lyrae $v_t$
and $v_r$ each make two, as in a reference disc (the model of the
$\feh<\ltFehA$ giants put through the same steps): the LMC's stars
rotate coherently at every metallicity.

To measure rotation and dispersion, we fit each metallicity bin of the
giants and the RR Lyrae on its own with a thin, inclined
disc at \ltDist\,kpc, as in the thin-disc models of
\citet{vanderMarel2002} and \citet{Vasiliev2018}, but fitted star by
star and with the perspective of each star's line of sight treated
exactly rather than to first order \citep[cf.][]{Vasiliev2020}.
In cylindrical coordinates $(R,\phi,z)$ in the disc frame, $z$ along
its spin, the mean velocity is the systemic motion plus a rotation curve
$v_{\rm rot}(R)$, rising as $\tanh$ and then linearly, and bisymmetric
in-plane streaming; about it the velocities follow a Gaussian
ellipsoid aligned with the disc, with dispersions $\sigma_R$,
$\sigma_\phi$ and $\sigma_z$, convolved with each star's errors, plus
an outlier component. The line-of-sight velocities mostly measure
$\sigma_z$ and the proper motions the in-plane dispersions
\citep{Vasiliev2018}, so the RR
Lyrae give $\sigma_R$ but not $\sigma_z$. We quote values at
$R=\ltRSig$\,kpc, the middle of the fitted radii; the RR Lyrae are
fitted at $\ltRrlFitIn^\circ<\rho<\ltRingOut^\circ$, beyond the crowded
inner disc. The posteriors, the sampler and the tests on mock data are in the
online material.

Fig.~\ref{fig:sims_kin} gives the main result of this Letter: the
rotational support of the LMC's stars against metallicity, compared
with that of simulated analogues introduced below. At every
metallicity the LMC's giants rotate fast and are kinematically cool.
From the most metal-poor bin to the most metal-rich, $v_{\rm rot}$
rises only from \ltKinVrotMP\ to \ltKinVrotMR\kms, while $\sigma_z$
falls from \ltKinSigMP\ to \ltKinSigMR\kms, so $v_{\rm rot}/\sigma_z$
grows from \ltKinRatioMP\ to \ltKinRatioMR. The last value is a lower
limit, because the less precise BOSS metallicities
scatter hotter, more metal-poor stars into the most metal-rich bin; APOGEE alone
gives \ltKinRatioMRApo. Even the giants
with $\feh<\ltFehA$ (median $\ltMPFehMed$, \ltMPPctLow\ per cent below
$\ltMPFehLow$), which formed with the first \ltFirstFrac\ per cent of
the LMC's stellar mass, are supported by rotation. Within this bin, the
\ltLowN\ giants with $\feh<\ltLowFeh$, which formed while the LMC had
built $\sim\ltLowMass$\Msun\ of stars, still have
$v_{\rm rot}/\sigma_z\approx\ltLowRatioLo$--\ltLowRatioHi. BOSS errors
scatter stars from $\feh\approx\ltLowInterLo$ to $\ltLowInterHi$ into this group
(\ltLowPurity\ per cent of those also observed by APOGEE lie below
$\ltLowFeh$), but in mock samples a non-rotating metal-poor population
falls below the measured value unless it makes up less than
\ltLowPurityMin\ per cent of the group (online material). The RR Lyrae give a consistent
$v_{\rm rot}/\sigma_R=\ltKinRrlRatioR$, against \ltKinRatioRMP\ for the
most metal-poor giants. The modelling also measures the size of each
population: between \ltHFitLo\ and \ltHFitHi\,kpc the selection-weighted
surface density of the giants falls exponentially, with a scale length
that grows from \ltHMR\,kpc for $\feh>\ltFehD$ to \ltHMP\,kpc for
$\feh<\ltFehA$ (\ltHRrl\,kpc for the RR Lyrae; half-mass radii of
\ltReMR, \ltReMP\ and \ltReRrl\,kpc if the exponential continued inside
\ltHFitLo\,kpc, where the bar dominates). These are present-day sizes:
heating and radial migration since the stars formed may have made the
old disc larger than it was at birth.

What matters for disc formation is the mass the LMC had when these
stars were born. Below $\feh=\ltFehA$ the LMC formed
$\ltEarlyMass$\Msun\ of stars (16th--84th percentiles
\ltEarlyMassLo--$\ltEarlyMassHi\times10^{\ltEarlyExp}$\Msun;
Fig.~\ref{fig:data}b), and below $\feh=\ltLowFeh$ only $\sim\ltLowMass$\Msun.
These are upper limits on its stellar mass when its most
metal-poor giants formed. The first is about \ltEarlyRatio\ times less than the
$2\times10^9$\Msun\ at which a simulated Milky Way progenitor forms its
disc, even with a resolved cold ISM \citep{Semenov2025a}, and
\ltEarlyRatioMW\ times less than the Milky Way had formed when it spun
up its disc \citep{Belokurov2023,Kurbatov2024}. The LMC built a lasting
stellar disc at a mass where most simulated galaxies are supported by
random motions \citep{ElBadry2018,Dekel2020}; in the Marvelous Massive
Dwarfs, discs form at lower masses, but from the stars born after the
merger that sets them up \citep{Geda2026}.

\begin{figure*}
\includegraphics[width=\textwidth]{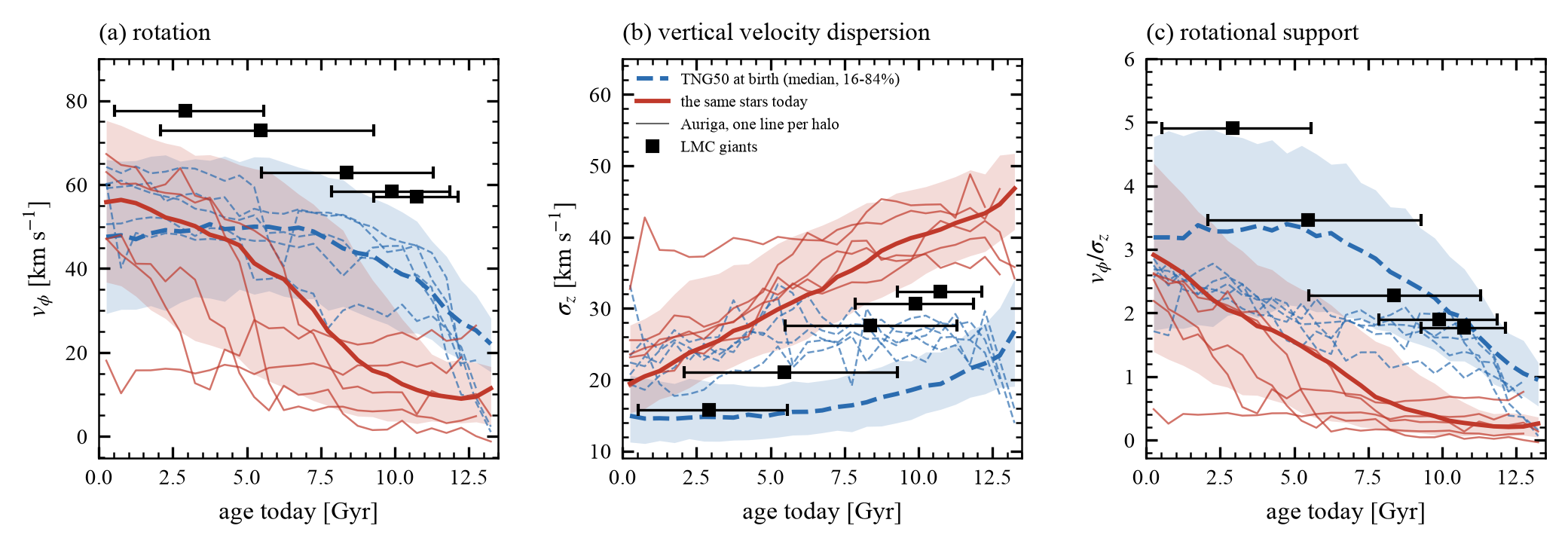}
\caption{The rotational support of the stars of simulated LMC analogues
  at birth (blue, dashed) and today (red, solid), against age, in
  cohorts of \ltCohort\,Gyr: (a) $v_\phi$, (b) $\sigma_z$, (c)
  $v_\phi/\sigma_z$. TNG50: median and 16th--84th percentiles over
  \ltNTngBirth\ analogues; Auriga: one thin line per galaxy. Squares:
  the LMC bins at their mean ages (bars: 16th--84th percentiles). The
  simulated values are intrinsic, at
  $\ltRMinBirth<R<\ltRBirth$\,kpc at birth and within \ltRToday\,kpc
  today; the LMC's are measured at $R=\ltRSig$\,kpc. The simulated
  stars are born with as much rotational support as the LMC's old stars
  have today, and the oldest of them have lost most of it since.
}
\label{fig:sims_birth}
\end{figure*}

We compare the LMC with all \ltNTng\ central galaxies of the TNG50
volume \citep{Nelson2019TNG50,Pillepich2019,Nelson2019release} with
stellar masses of $\ltMstarLo$--$\ltMstarHi\times10^9$\Msun\ and total
masses of $\ltMtotLo$--$\ltMtotHi$\Msun, and with the zoom-in
simulations Auriga \citep[\ltNAu\ dwarfs;][]{Grand2024} and FIRE-2
\citep[\ltNFire\ galaxies;][]{Wetzel2023}, whose mass resolution is
about ten times finer. TNG50 and Auriga treat star-forming gas as a pressurised
multiphase medium \citep{Springel2003}; FIRE-2 resolves its cold phase
and forms stars in bursts at these masses. Each analogue is observed with the LMC's samples and errors
and fitted with the same model, and its metallicities are matched to
the LMC's by rank in formed mass (details in the online material).

None of the suites reproduces the LMC's early disc, and only TNG50
reproduces its later one. In the three most metal-rich bins,
\ltTngPctC, \ltTngPctD\ and \ltTngPctE\ per cent of the TNG50 analogues
reach the LMC's $v_{\rm rot}/\sigma_z$. The Auriga and FIRE-2 analogues
fall short at every metallicity: even for their most metal-rich stars
the highest $v_{\rm rot}/\sigma_z$ is \ltAuRatioMRMax\ and
\ltFireRatioMRMax, against the LMC's \ltKinRatioMR--\ltKinRatioMRApo,
although these suites contain only \ltNAu\ and \ltNFire\ galaxies. In the most metal-poor bin no
TNG50 analogue reaches the LMC either: the median analogue has
$v_{\rm rot}/\sigma_z=\ltTngRatioMP$ and the highest \ltTngRatioMPMax,
against the LMC's \ltKinRatioMP; the only analogue as
cold as the LMC barely rotates ($v_{\rm rot}=\ltTngColdVrot$\kms).

\section{The absence of early stellar discs in simulated LMC analogues}\label{sec:sims}

\begin{figure*}
\includegraphics[width=\textwidth]{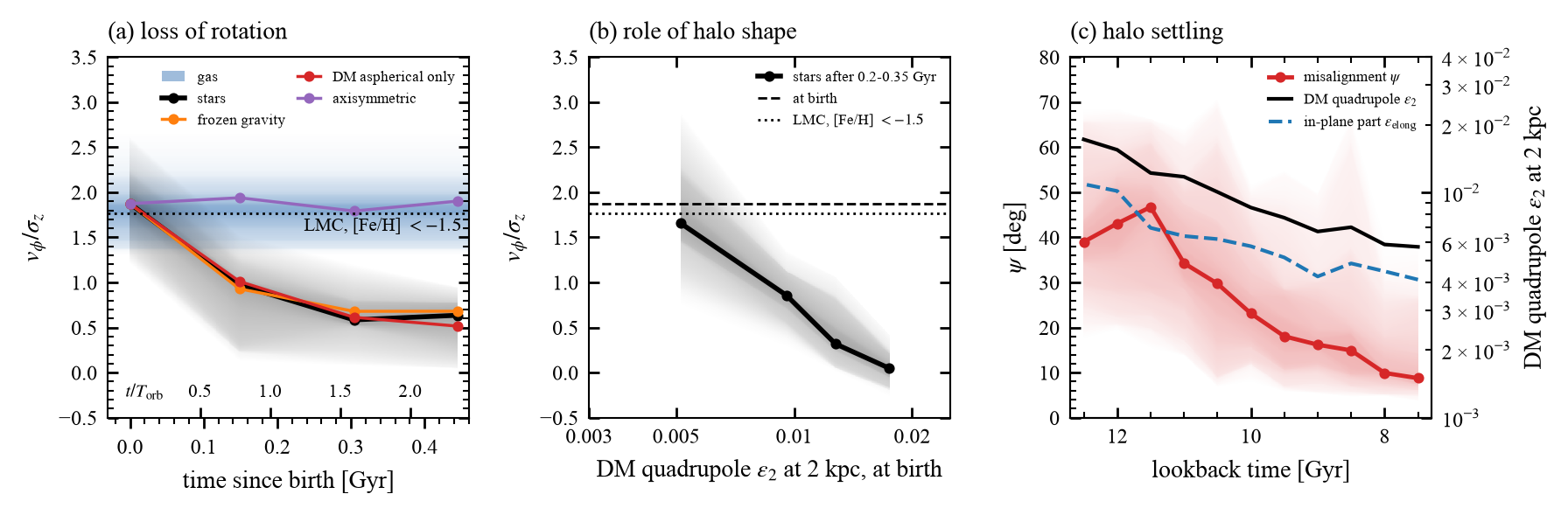}
\caption{How the early TNG50 cohorts lose their rotation, and why. (a)
  $v_\phi/\sigma_z$ about each cohort's spin at birth against time
  since birth (inner labels: orbital periods at \ltROrb\,kpc) for the
  real stars (black) and for test particles in the progenitor's gravity
  frozen at birth (orange), with only the dark matter aspherical (red)
  and made axisymmetric about the cohort's spin (purple); blue band:
  the star-forming gas; dotted: the LMC's $\feh<\ltFehA$ giants (at
  $R=\ltRSig$\,kpc, for reference). (b) $v_\phi/\sigma_z$ after
  \ltTAfterA--\ltTAfterB\,Gyr against the dark-matter quadrupole
  $\epsilon_2$ at birth; dashed: at birth. (c) Misalignment $\psi$ of the
  halo's minor axis with the disc spin (red, left), and $\epsilon_2$
  (black) with its in-plane part (blue dashed; right), against lookback
  time. The rotation is halved within one orbital period, and mostly
  lost within two, because the halo is aspherical and misaligned with
  the disc; it settles only after the LMC's oldest stars formed.
}
\label{fig:scramble}
\end{figure*}

On a closer look, at birth the stars of the simulated analogues are
supported by rotation; however, they lose this support quickly. TNG50 and Auriga
record the position and velocity of every star particle at birth, so
the analogues' stars can be measured when they formed and today. We
measure the early discs at $\ltRMinBirth<R<\ltRBirth$\,kpc in their
progenitors, where TNG50 resolves them (at 3--5\,kpc it has
\ltRadPartRatio\ times as many particles) and where most old stars seen
today at the LMC's radius formed (only \ltRadBornOuter\ per cent of
them were born at 3--5\,kpc); the LMC is only a
reference in
Figs~\ref{fig:sims_birth} and~\ref{fig:scramble}, whereas
Fig.~\ref{fig:sims_kin} compares at the same radius.

Fig.~\ref{fig:sims_birth} compares the rotational support of the stars
of the simulated analogues at birth and at $z=0$: the mass-weighted
mean azimuthal velocity $v_\phi$ and vertical dispersion $\sigma_z$ of
each cohort, at birth relative to its progenitor and about its own
spin, today in the disc frame. The LMC bins are placed at the ages that
the age--metallicity relation of \citet{Povick2024} gives them. At
birth even the oldest simulated stars rotate: the cohorts
\ltOldAgeLo--\ltOldAgeHi\,Gyr old formed with
$v_\phi/\sigma_z=\ltTngRbMed$ in TNG50 and \ltAuRbMed\ in Auriga, as
high as the LMC's \ltKinRatioMP\ and \ltKinRatioB\ today. The same
stars today have $v_\phi/\sigma_z=\ltTngRtMed$ in TNG50 and
\ltAuRtMed\ in Auriga: their rotation has fallen from \ltTngVbMed\ to
\ltTngVtMed\kms\ and their $\sigma_z$ has risen from \ltTngSbMed\ to
\ltTngStMed\kms.

To find how fast the rotation is lost and why, we follow TNG50
progenitors back in time (Fig.~\ref{fig:scramble}): \ltNCohorts\ cohorts born
at $z=\ltZSnapA$--\ltZSnapB\ in \ltNCohortGal\ of \ltNFull\ progenitors
with outputs at every snapshot, and the dark matter of a larger sample
of \ltNSample\ that includes them. Here a cohort is the stars formed in the
\ltWindow\,Myr before an output (its birth snapshot), followed for
\ltDtMax\,Gyr; its $v_\phi/\sigma_z$ is measured about its spin at
birth. The same stars are also moved as test particles through their
progenitor's gravity \citep[computed with AGAMA;][]{Vasiliev2019}:
frozen at birth, with only the dark matter aspherical, and made
axisymmetric about the cohort's spin.

Within \ltScrTSecond\,Gyr, less than one orbital period
($T_{\rm orb}=\ltTOrb$\,Gyr at \ltROrb\,kpc), the $v_\phi/\sigma_z$ of
the early cohorts falls from \ltScrStart\ to \ltScrSecond, and after
\ltScrTEnd\,Gyr it is \ltScrEnd, while the star-forming gas keeps
$v_\phi/\sigma_z\approx\ltGasRatio$ about its own, possibly changing,
spin. Test particles lose their rotation as fast in the gravity frozen
at birth (\ltFrozenEnd\ at the end) and when only the dark matter is
aspherical (\ltDmEnd), but keep it when the gravity is axisymmetric
about their spin (\ltAxiEnd). The rotation is therefore destroyed by the
shape of the halo in which the stars are born, without later mergers,
and by the part of its asphericity that is not axisymmetric about the
disc's spin.

Fig.~\ref{fig:scramble}(b) relates the loss to this part of the halo
at birth. We measure it with $\epsilon_2$, the amplitude of the
quadrupole of the dark-matter potential at $r=\ltSetRZero$\,kpc that is
not axisymmetric about the disc's spin, relative to the monopole: it
grows with a tilt of the halo relative to the disc and with its
elongation in the disc plane, but not with a flattening along the
spin (definition in the online material). The rotation that survives
falls steadily with $\epsilon_2$ (Spearman rank correlation
\ltEpsSpearman\ over \ltNCohortsB\ cohorts): cohorts born at
$\epsilon_2\approx\ltEpsLow$ keep $v_\phi/\sigma_z\approx\ltVsEpsLow$,
close to their value at birth, while those born at
$\epsilon_2\approx\ltEpsHigh$ end with \ltVsEpsHigh. An early disc
survives only in a halo nearly axisymmetric about its spin.

Fig.~\ref{fig:scramble}(c) shows how the halo changes. At a lookback
time of \ltLbEarly\,Gyr its minor axis is misaligned with the disc
spin by $\psi\approx\ltPsiEarly^\circ$ and $\epsilon_2\approx\ltEpsEarly$,
the value at which cohorts lose all their rotation. On a timescale of \ltSettleLo--\ltSettleHi\,Gyr the halo
settles: its minor axis turns towards the disc's spin, its elongation
in the disc plane shrinks, and it becomes somewhat rounder (the
minor-to-major axis ratio of the dark matter within \ltSetRIn\,kpc
grows from $c/a=\ltCaEarly$ to \ltCaLate). By \ltLbLate\,Gyr ago
$\psi\approx\ltPsiLate^\circ$ and $\epsilon_2\approx\ltEpsLate$, close to
where cohorts keep their rotation: the halo becomes quiet only after
the LMC's oldest stars formed, \ltOldAgeLo--\ltOldAgeHi\,Gyr ago.

In Auriga the experiment of
Fig.~\ref{fig:scramble}(a), run on the early cohorts with the gravity
frozen at birth, gives the same result (online material). In FIRE-2
the oldest stars also lose their rotation quickly,
but they are born with less of it, $v_\phi/\sigma_z=\ltFireRbMed$ for the
cohorts \ltOldAgeLo--\ltOldAgeHi\,Gyr old, as in the FIREbox volume,
where dwarfs below $M_\star\approx10^9$\Msun\ form stars from gas too
turbulent to settle into a disc \citep{Benavides2025}. We have not
tested what scrambles the orbits in FIRE-2.

\section{Discussion and conclusions}\label{sec:discussion}

\begin{figure*}
\includegraphics[width=\textwidth]{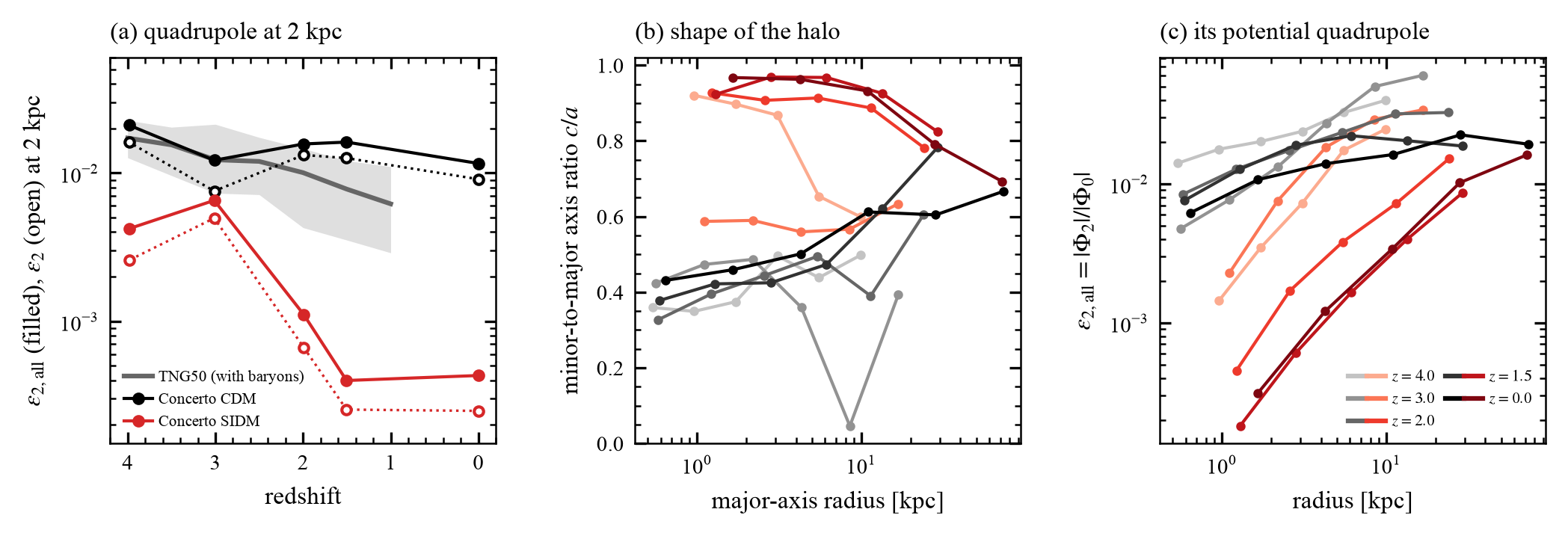}
\caption{The centre of an LMC-mass halo in cold (CDM) and
  self-interacting (SIDM) dark matter (SIDM Concerto Halo104, dark
  matter only). (a) Quadrupole at \ltREps\,kpc against redshift: CDM
  (black) and SIDM (red), the full quadrupole including its
  axisymmetric part ($\epsilon_{2,\rm all}$, filled) and the part not
  axisymmetric about the halo's minor axis ($\epsilon_2$, open); grey: TNG50. (b) Minor-to-major axis ratio
  $c/a$ and (c) $\epsilon_{2,\rm all}$ against radius, light to dark
  from $z=\ltSiZFirst$ to 0. With self-interactions the centre of the
  halo is round already at $z\approx\ltSiZFirst$; in CDM it stays
  flattened and aspherical.
}
\label{fig:sidm}
\end{figure*}

\subsection{Halo roundness in SIDM and CDM}\label{sec:sidm}

Collisions between dark-matter particles could make the centre of the
halo rounder, because scattering exchanges
energy and angular momentum and isotropises the orbits in the dense
inner halo \citep{Peter2013,Vargya2022}. Fig.~\ref{fig:sidm} tests this
with the CDM and self-interacting dark matter (SIDM) runs of the
LMC-mass halo Halo104 of SIDM Concerto \citep{Nadler2025}, from the
same initial conditions, with dark matter only and a velocity-dependent
cross-section per unit mass of normalisation
$\sigma_0/m=\ltSigmaZero$\,cm$^2$\,g$^{-1}$ that falls as $v^{-4}$
above $w=\ltWSidm$\kms.

In CDM the non-axisymmetric part of the quadrupole at \ltREps\,kpc
about the halo's minor axis is $\epsilon_2\approx\ltSiCdmNaLo$--\ltSiCdmNaHi\
at all redshifts, similar to the TNG50 progenitors. In SIDM the full
quadrupole, including its axisymmetric part, is already small at
$z=\ltSiZFirst$, $\epsilon_{2,\rm all}\approx\ltSiEpsFirst$, and falls to
$\approx\ltSiEpsLate$ by $z=1.5$, about \ltSiRatioLate\ times below CDM.
As the full quadrupole is the same about any axis and bounds its
non-axisymmetric part about every axis, a disc of any orientation in
the SIDM halo would feel $\epsilon_2\lesssim\ltSiEpsFirst$ at
$z=\ltSiZFirst$, at or below the value at which the TNG50 cohorts keep
their rotation. The SIDM halo is nearly round within $\sim10$\,kpc
($c/a\approx\ltSiCaSidmMed$ at \ltREps\,kpc, against
\ltSiCaCdmLo--\ltSiCaCdmHi\ in CDM), where the early disc's stars
orbit.

The cross-section falls with collision speed, to
$\approx0.03$\,cm$^2$\,g$^{-1}$ at the $\sim1000$\kms\ speeds in galaxy
clusters, below the upper limits that clusters set, so the model is not
excluded by them (\citealt{Nadler2025} and references therein).
A baryonic disc flattens an SIDM halo along the disc and aligns it
with the disc more closely than a CDM halo \citep{Sameie2018}; this keeps
the halo axisymmetric about the disc's spin, the condition under which
the disc survives (Fig.~\ref{fig:scramble}b). Milky Way-mass SIDM haloes
with baryons are no rounder than CDM ones \citep{Vargya2022}, but their
alignment with the disc was not measured. Whether SIDM preserves an early LMC disc
remains to be tested with hydrodynamical SIDM simulations.

\subsection{Conclusions}\label{sec:conclusions}

\begin{enumerate}
\item The LMC's oldest stars are in a rotating disc: even its giants
  with $\feh<\ltFehA$ have $v_{\rm rot}/\sigma_z\approx\ltKinRatioMP$,
  and the rotational support grows with metallicity, to
  $\gtrsim\ltKinRatioMR$. The RR Lyrae are consistent with this.
  The disc reaches back to the LMC's first $\sim\ltLowMass$\Msun\ of
  stars: its giants with $\feh<\ltLowFeh$ have
  $v_{\rm rot}/\sigma_z\approx\ltLowRatioLo$--\ltLowRatioHi. Measurement
  errors mix more metal-rich stars into this group, but their rotation
  alone could explain the measured value only if they made up more than
  \ltLowInterMax\ per cent of it.
\item The LMC had this disc when it had formed at most
  $\ltEarlyMass$\Msun\ of stars, about \ltEarlyRatio\ times less than
  the mass at which simulated Milky Way progenitors form their discs:
  low-mass galaxies can build lasting stellar discs far below the
  threshold seen in most simulations.
\item None of the \ltNTng\ TNG50, \ltNAu\ Auriga and \ltNFire\ FIRE-2
  LMC analogues, observed with the LMC's samples and errors, matches
  the rotational support of its most metal-poor stars; only TNG50
  reproduces its metal-rich disc.
\item In TNG50 and Auriga the oldest stars are born in rotating discs
  but halve their rotation within one orbital period and lose most of
  it within two, scrambled by the aspherical, misaligned centre of the
  dark-matter halo, which settles only after the LMC's oldest stars
  formed. In FIRE-2 they are born with less rotation.
\item In an SIDM resimulation of an LMC-mass halo, the inner halo is
  nearly round already at $z\approx\ltSiZFirst$. Self-interactions are
  one way in which early discs could survive, but the cross-section is
  large and the run has no baryons; changes to star formation and
  feedback may offer others.
\end{enumerate}

4MOST \citep{Cioni2019} will extend the measurement to the LMC's most
metal-poor stars. Line-of-sight velocities of red giants in other
nearby galaxies already show both outcomes: in M33, as in the LMC,
most old giants corotate with the disc
\citep{Gilbert2022,Cullinane2023}, whereas in NGC~6822 and WLM the
metal-poor giants are hotter than the younger ones
\citep{Leaman2009,Swan2016}. Larger samples with ages will show where
the earliest stellar discs survived.

\section*{Acknowledgements}
We thank Vincenzo Ripepi and Marcella Marconi for the discussion of
RR Lyrae, at the Rubin LSST-Europe-8 conference in Budapest in 2026,
that started this work. We thank Ethan Nadler and Demao Kong for sharing the LMC-mass SIDM
Concerto simulation. We thank Andrey Kravtsov for useful comments that
helped to improve the manuscript. The analysis code and parts of the text were prepared
with the help of AI tools (Claude, Anthropic; and, at some stages, Codex,
OpenAI); the author checked all results and takes full responsibility for the
content.

This work has made use of data from the European Space Agency (ESA) mission
{\it Gaia} (\url{https://www.cosmos.esa.int/gaia}), processed by the {\it Gaia}
Data Processing and Analysis Consortium (DPAC,
\url{https://www.cosmos.esa.int/web/gaia/dpac/consortium}). Funding for the DPAC
has been provided by national institutions, in particular the institutions
participating in the {\it Gaia} Multilateral Agreement.

Funding for the Sloan Digital Sky Survey IV has been provided by the Alfred P. Sloan Foundation, the U.S. Department of Energy Office of Science, and the Participating Institutions. SDSS-IV acknowledges support and resources from the Center for High Performance Computing at the University of Utah. The SDSS website is \href{https://www.sdss4.org}{www.sdss4.org}. SDSS-IV is managed by the Astrophysical Research Consortium for the Participating Institutions of the SDSS Collaboration including the Brazilian Participation Group, the Carnegie Institution for Science, Carnegie Mellon University, Center for Astrophysics $|$ Harvard \& Smithsonian, the Chilean Participation Group, the French Participation Group, Instituto de Astrof\'isica de Canarias, The Johns Hopkins University, Kavli Institute for the Physics and Mathematics of the Universe (IPMU) / University of Tokyo, the Korean Participation Group, Lawrence Berkeley National Laboratory, Leibniz Institut f\"ur Astrophysik Potsdam (AIP), Max-Planck-Institut f\"ur Astronomie (MPIA Heidelberg), Max-Planck-Institut f\"ur Astrophysik (MPA Garching), Max-Planck-Institut f\"ur Extraterrestrische Physik (MPE), National Astronomical Observatories of China, New Mexico State University, New York University, University of Notre Dame, Observat\'ario Nacional / MCTI, The Ohio State University, Pennsylvania State University, Shanghai Astronomical Observatory, United Kingdom Participation Group, Universidad Nacional Aut\'onoma de M\'exico, University of Arizona, University of Colorado Boulder, University of Oxford, University of Portsmouth, University of Utah, University of Virginia, University of Washington, University of Wisconsin, Vanderbilt University, and Yale University. APOGEE DR17 \citep{Abdurrouf2022} is part of SDSS-IV \citep{Blanton2017,Majewski2017,Wilson2019}.

Funding for the Sloan Digital Sky Survey V has been provided by the Alfred P. Sloan Foundation, the Heising-Simons Foundation, the National Science Foundation, and the Participating Institutions. SDSS acknowledges support and resources from the Center for High-Performance Computing at the University of Utah. SDSS telescopes are located at Apache Point Observatory, funded by the Astrophysical Research Consortium and operated by New Mexico State University, and at Las Campanas Observatory, operated by the Carnegie Institution for Science. The SDSS web site is \href{https://www.sdss.org}{www.sdss.org}. SDSS is managed by the Astrophysical Research Consortium for the Participating Institutions of the SDSS Collaboration, including the Carnegie Institution for Science, Chilean National Time Allocation Committee (CNTAC) ratified researchers, Caltech, the Gotham Participation Group, Harvard University, Heidelberg University, The Flatiron Institute, The Johns Hopkins University, L'Ecole polytechnique f\'{e}d\'{e}rale de Lausanne (EPFL), Leibniz-Institut f\"{u}r Astrophysik Potsdam (AIP), Max-Planck-Institut f\"{u}r Astronomie (MPIA Heidelberg), Max-Planck-Institut f\"{u}r Extraterrestrische Physik (MPE), Nanjing University, National Astronomical Observatories of China (NAOC), New Mexico State University, The Ohio State University, Pennsylvania State University, Smithsonian Astrophysical Observatory, Space Telescope Science Institute (STScI), the Stellar Astrophysics Participation Group, Universidad Nacional Aut\'{o}noma de M\'{e}xico, University of Arizona, University of Colorado Boulder, University of Illinois at Urbana-Champaign, University of Toronto, University of Utah, University of Virginia, Yale University, and Yunnan University. The BOSS spectra were taken with the 2.5-m Sloan \citep{Gunn2006} and du Pont \citep{Bowen1973} telescopes \citep{Smee2013}.

This publication makes use of data products from the Two Micron All Sky Survey,
which is a joint project of the University of Massachusetts and the Infrared
Processing and Analysis Center/California Institute of Technology, funded by the
National Aeronautics and Space Administration and the National Science
Foundation.

The IllustrisTNG simulations were undertaken with compute time awarded by the
Gauss Centre for Supercomputing (GCS) under GCS Large-Scale Projects GCS-ILLU and
GCS-DWAR on the GCS share of the supercomputer Hazel Hen at the High Performance
Computing Center Stuttgart (HLRS), as well as on the machines of the Max Planck
Computing and Data Facility (MPCDF) in Garching, Germany. We have used
simulations from the Auriga Project \citep{Grand2017} public data release
\citep{Grand2024} available at \url{https://wwwmpa.mpa-garching.mpg.de/auriga/data.html}.
We use the publicly-available FIRE-2 cosmological zoom-in simulations
\citep{Wetzel2023,Wetzel2025}, from the Feedback In Realistic Environments (FIRE)
project, generated using the Gizmo code \citep{Hopkins2015} and the FIRE-2
physics model \citep{Hopkins2018}; the runs m11d, m11e and m11h were introduced
by \citet{ElBadry2018}.

This paper made use of the Whole Sky Database (wsdb) created and maintained by
Sergey Koposov at the Institute of Astronomy, Cambridge with financial support
from the Science \& Technology Facilities Council (STFC) and the European
Research Council (ERC). This publication made use of the Q3C software
\citep{Koposov2006}. This research has made use of the VizieR catalogue access
tool, CDS, Strasbourg, France (DOI: 10.26093/cds/vizier); the original
description of the VizieR service was published in \citet{Ochsenbein2000}. This
research has made use of the SIMBAD database, operated at CDS, Strasbourg,
France \citep{Wenger2000}.

This work made use of Astropy (\url{https://www.astropy.org}), a
community-developed core Python package and an ecosystem of tools and resources
for astronomy \citep{Astropy2013,Astropy2018,Astropy2022}. Some of the results in
this paper have been derived using the healpy and HEALPix package
\citep{Zonca2019,Gorski2005}. We also used NumPy \citep{Harris2020}, SciPy
\citep{Virtanen2020}, Matplotlib \citep{Hunter2007}, pandas \citep{McKinney2010},
astroquery \citep{Ginsburg2019}, JAX \citep{Bradbury2018}, NumPyro
\citep{Bingham2018,Phan2019}, corner \citep{ForemanMackey2016}, gala
\citep{PriceWhelan2017}, dustmaps \citep{Green2018}, AGAMA \citep{Vasiliev2019},
h5py (\url{https://www.h5py.org}), PyArrow (\url{https://arrow.apache.org}), PyVO
(\url{https://pyvo.readthedocs.io}) and ezpadova
(\url{https://github.com/mfouesneau/ezpadova}).

\section*{Data availability}
All observational data used in this Letter are public. {\it Gaia} DR3 is
available from the {\it Gaia} archive (\url{https://gea.esac.esa.int/archive/}),
APOGEE DR17 and the SDSS-V DR20 BOSS spectra from the SDSS
(\url{https://www.sdss.org}), and the OGLE-IV RR Lyrae catalogues from the OGLE
archive (\url{https://ogle.astrouw.edu.pl}). The TNG50, Auriga and FIRE-2
simulations are public \citep{Nelson2019release,Grand2024,Wetzel2023}. This work used data from SIDM Concerto \citep{Nadler2025}. The SIDM Concerto
data is publicly available on Zenodo at DOI 10.5281/zenodo.14933624
\citep{Nadler2025data}. The derived catalogues (the cross-matched giant
and RR Lyrae samples with their selection weights) and the posterior samples of
the kinematic fits are at \url{https://doi.org/10.5281/zenodo.24447285}. The code that
makes every figure, table and number of this Letter is at
\url{https://github.com/vasilybelokurov/lmc-ancient-disc}. It includes a list
of every input with its origin and checksum (\texttt{data/sources.yaml}) and a
website with a notebook for each figure
(\url{https://vasilybelokurov.github.io/lmc-ancient-disc/}).

\bibliographystyle{mnras}
\bibliography{refs}

\begin{thebibliography}{}
\makeatletter
\relax
\def\mn@urlcharsother{\let\do\@makeother \do\$\do\&\do\#\do\^\do\_\do\%\do\~}
\def\mn@doi{\begingroup\mn@urlcharsother \@ifnextchar [ {\mn@doi@}
  {\mn@doi@[]}}
\def\mn@doi@[#1]#2{\def\@tempa{#1}\ifx\@tempa\@empty \href
  {http://dx.doi.org/#2} {doi:#2}\else \href {http://dx.doi.org/#2} {#1}\fi
  \endgroup}
\def\mn@eprint#1#2{\mn@eprint@#1:#2::\@nil}
\def\mn@eprint@arXiv#1{\href {http://arxiv.org/abs/#1} {{\tt arXiv:#1}}}
\def\mn@eprint@dblp#1{\href {http://dblp.uni-trier.de/rec/bibtex/#1.xml}
  {dblp:#1}}
\def\mn@eprint@#1:#2:#3:#4\@nil{\def\@tempa {#1}\def\@tempb {#2}\def\@tempc
  {#3}\ifx \@tempc \@empty \let \@tempc \@tempb \let \@tempb \@tempa \fi \ifx
  \@tempb \@empty \def\@tempb {arXiv}\fi \@ifundefined
  {mn@eprint@\@tempb}{\@tempb:\@tempc}{\expandafter \expandafter \csname
  mn@eprint@\@tempb\endcsname \expandafter{\@tempc}}}

\bibitem[\protect\citeauthoryear{{Abdurro'uf} et~al.,}{{Abdurro'uf}
  et~al.}{2022}]{Abdurrouf2022}
{Abdurro'uf} et~al., 2022, \mn@doi [\apjs] {10.3847/1538-4365/ac4414}, \href
  {https://ui.adsabs.harvard.edu/abs/2022ApJS..259...35A} {259, 35}

\bibitem[\protect\citeauthoryear{{Alcock} et~al.,}{{Alcock}
  et~al.}{1996}]{Alcock1996}
{Alcock} C.,  et~al., 1996, \mn@doi [\aj] {10.1086/117859}, \href
  {https://scixplorer.org/abs/1996AJ....111.1146A} {111, 1146}

\bibitem[\protect\citeauthoryear{{Astropy Collaboration} et~al.,}{{Astropy
  Collaboration} et~al.}{2013}]{Astropy2013}
{Astropy Collaboration} et~al., 2013, \mn@doi [\aap]
  {10.1051/0004-6361/201322068}, \href
  {https://ui.adsabs.harvard.edu/abs/2013A&A...558A..33A} {558, A33}

\bibitem[\protect\citeauthoryear{{Astropy Collaboration} et~al.,}{{Astropy
  Collaboration} et~al.}{2018}]{Astropy2018}
{Astropy Collaboration} et~al., 2018, \mn@doi [\aj] {10.3847/1538-3881/aabc4f},
  \href {https://ui.adsabs.harvard.edu/abs/2018AJ....156..123A} {156, 123}

\bibitem[\protect\citeauthoryear{{Astropy Collaboration} et~al.,}{{Astropy
  Collaboration} et~al.}{2022}]{Astropy2022}
{Astropy Collaboration} et~al., 2022, \mn@doi [\apj]
  {10.3847/1538-4357/ac7c74}, \href
  {https://ui.adsabs.harvard.edu/abs/2022ApJ...935..167A} {935, 167}

\bibitem[\protect\citeauthoryear{{Belokurov} \& {Koposov}}{{Belokurov} \&
  {Koposov}}{2016}]{Belokurov2016}
{Belokurov} V.,  {Koposov} S.~E.,  2016, \mn@doi [\mnras]
  {10.1093/mnras/stv2688}, \href
  {https://scixplorer.org/abs/2016MNRAS.456..602B} {456, 602}

\bibitem[\protect\citeauthoryear{{Belokurov} \& {Kravtsov}}{{Belokurov} \&
  {Kravtsov}}{2022}]{Belokurov2022}
{Belokurov} V.,  {Kravtsov} A.,  2022, \mn@doi [\mnras]
  {10.1093/mnras/stac1267}, \href
  {https://scixplorer.org/abs/2022MNRAS.514..689B} {514, 689}

\bibitem[\protect\citeauthoryear{{Belokurov} \& {Kravtsov}}{{Belokurov} \&
  {Kravtsov}}{2023}]{Belokurov2023}
{Belokurov} V.,  {Kravtsov} A.,  2023, \mn@doi [\mnras]
  {10.1093/mnras/stad2241}, \href
  {https://scixplorer.org/abs/2023MNRAS.525.4456B} {525, 4456}

\bibitem[\protect\citeauthoryear{{Belokurov}, {Erkal}, {Deason}, {Koposov}, {De
  Angeli}, {Evans}, {Fraternali}  \& {Mackey}}{{Belokurov}
  et~al.}{2017}]{Belokurov2017}
{Belokurov} V.,  {Erkal} D.,  {Deason} A.~J.,  {Koposov} S.~E.,  {De Angeli}
  F.,  {Evans} D.~W.,  {Fraternali} F.,   {Mackey} D.,  2017, \mn@doi [\mnras]
  {10.1093/mnras/stw3357}, \href
  {https://scixplorer.org/abs/2017MNRAS.466.4711B} {466, 4711}

\bibitem[\protect\citeauthoryear{{Belokurov}, {Sanders}, {Fattahi}, {Smith},
  {Deason}, {Evans}  \& {Grand}}{{Belokurov} et~al.}{2020}]{Belokurov2020}
{Belokurov} V.,  {Sanders} J.~L.,  {Fattahi} A.,  {Smith} M.~C.,  {Deason}
  A.~J.,  {Evans} N.~W.,   {Grand} R. J.~J.,  2020, \mn@doi [\mnras]
  {10.1093/mnras/staa876}, \href
  {https://scixplorer.org/abs/2020MNRAS.494.3880B} {494, 3880}

\bibitem[\protect\citeauthoryear{{Benavides} et~al.,}{{Benavides}
  et~al.}{2025}]{Benavides2025}
{Benavides} J.~A.,  et~al., 2025, \mn@doi [\mnras] {10.1093/mnras/staf1847},
  \href {https://ui.adsabs.harvard.edu/abs/2025MNRAS.544.4651B} {544, 4651}

\bibitem[\protect\citeauthoryear{{Bingham} et~al.,}{{Bingham}
  et~al.}{2018}]{Bingham2018}
{Bingham} E.,  et~al., 2018, \mn@doi [arXiv e-prints]
  {10.48550/arXiv.1810.09538}, \href
  {https://ui.adsabs.harvard.edu/abs/2018arXiv181009538B} {p. arXiv:1810.09538}

\bibitem[\protect\citeauthoryear{{Blanton} et~al.,}{{Blanton}
  et~al.}{2017}]{Blanton2017}
{Blanton} M.~R.,  et~al., 2017, \mn@doi [\aj] {10.3847/1538-3881/aa7567}, \href
  {https://ui.adsabs.harvard.edu/abs/2017AJ....154...28B} {154, 28}

\bibitem[\protect\citeauthoryear{{Bonaca}, {Conroy}, {Wetzel}, {Hopkins}  \&
  {Kere{\v{s}}}}{{Bonaca} et~al.}{2017}]{Bonaca2017}
{Bonaca} A.,  {Conroy} C.,  {Wetzel} A.,  {Hopkins} P.~F.,   {Kere{\v{s}}} D.,
  2017, \mn@doi [\apj] {10.3847/1538-4357/aa7d0c}, \href
  {https://scixplorer.org/abs/2017ApJ...845..101B} {845, 101}

\bibitem[\protect\citeauthoryear{{Borissova}, {Minniti}, {Rejkuba}  \&
  {Alves}}{{Borissova} et~al.}{2006}]{Borissova2006}
{Borissova} J.,  {Minniti} D.,  {Rejkuba} M.,   {Alves} D.,  2006, \mn@doi
  [\aap] {10.1051/0004-6361:20054132}, \href
  {https://ui.adsabs.harvard.edu/abs/2006A&A...460..459B} {460, 459}

\bibitem[\protect\citeauthoryear{{Bowen} \& {Vaughan}}{{Bowen} \&
  {Vaughan}}{1973}]{Bowen1973}
{Bowen} I.~S.,  {Vaughan} Jr. A.~H.,  1973, \mn@doi [\ao]
  {10.1364/AO.12.001430}, \href
  {https://ui.adsabs.harvard.edu/abs/1973ApOpt..12.1430B} {12, 1430}

\bibitem[\protect\citeauthoryear{{Bradbury} et~al.,}{{Bradbury}
  et~al.}{2018}]{Bradbury2018}
{Bradbury} J.,  et~al., 2018, {JAX: composable transformations of Python+NumPy
  programs}, \url{http://github.com/jax-ml/jax}

\bibitem[\protect\citeauthoryear{{Bressan}, {Marigo}, {Girardi}, {Salasnich},
  {Dal Cero}, {Rubele}  \& {Nanni}}{{Bressan} et~al.}{2012}]{Bressan2012}
{Bressan} A.,  {Marigo} P.,  {Girardi} L.,  {Salasnich} B.,  {Dal Cero} C.,
  {Rubele} S.,   {Nanni} A.,  2012, \mn@doi [\mnras]
  {10.1111/j.1365-2966.2012.21948.x}, \href
  {https://ui.adsabs.harvard.edu/abs/2012MNRAS.427..127B} {427, 127}

\bibitem[\protect\citeauthoryear{{Carrera}, {Gallart}, {Aparicio}  \&
  {Hardy}}{{Carrera} et~al.}{2011}]{Carrera2011}
{Carrera} R.,  {Gallart} C.,  {Aparicio} A.,   {Hardy} E.,  2011, \mn@doi [\aj]
  {10.1088/0004-6256/142/2/61}, \href
  {https://scixplorer.org/abs/2011AJ....142...61C} {142, 61}

\bibitem[\protect\citeauthoryear{{Cioni} et~al.,}{{Cioni}
  et~al.}{2019}]{Cioni2019}
{Cioni} M.-R.~L.,  et~al., 2019, \mn@doi [The Messenger]
  {10.18727/0722-6691/5128}, \href
  {https://scixplorer.org/abs/2019Msngr.175...54C} {175, 54}

\bibitem[\protect\citeauthoryear{{Clauwens}, {Schaye}, {Franx}  \&
  {Bower}}{{Clauwens} et~al.}{2018}]{Clauwens2018}
{Clauwens} B.,  {Schaye} J.,  {Franx} M.,   {Bower} R.~G.,  2018, \mn@doi
  [\mnras] {10.1093/mnras/sty1229}, \href
  {https://scixplorer.org/abs/2018MNRAS.478.3994C} {478, 3994}

\bibitem[\protect\citeauthoryear{{Clementini} et~al.,}{{Clementini}
  et~al.}{2023}]{Clementini2023}
{Clementini} G.,  et~al., 2023, \mn@doi [\aap] {10.1051/0004-6361/202243964},
  \href {https://ui.adsabs.harvard.edu/abs/2023A&A...674A..18C} {674, A18}

\bibitem[\protect\citeauthoryear{{Conroy} et~al.,}{{Conroy}
  et~al.}{2022}]{Conroy2022}
{Conroy} C.,  et~al., 2022, \mn@doi [arXiv e-prints]
  {10.48550/arXiv.2204.02989}, \href
  {https://scixplorer.org/abs/2022arXiv220402989C} {p. arXiv:2204.02989}

\bibitem[\protect\citeauthoryear{{Cullinane}, {Mackey}, {Da Costa}, {Erkal},
  {Koposov}  \& {Belokurov}}{{Cullinane} et~al.}{2022a}]{Cullinane2022a}
{Cullinane} L.~R.,  {Mackey} A.~D.,  {Da Costa} G.~S.,  {Erkal} D.,  {Koposov}
  S.~E.,   {Belokurov} V.,  2022a, \mn@doi [\mnras] {10.1093/mnras/stab3350},
  \href {https://ui.adsabs.harvard.edu/abs/2022MNRAS.510..445C} {510, 445}

\bibitem[\protect\citeauthoryear{{Cullinane}, {Mackey}, {Da Costa}, {Erkal},
  {Koposov}  \& {Belokurov}}{{Cullinane} et~al.}{2022b}]{Cullinane2022b}
{Cullinane} L.~R.,  {Mackey} A.~D.,  {Da Costa} G.~S.,  {Erkal} D.,  {Koposov}
  S.~E.,   {Belokurov} V.,  2022b, \mn@doi [\mnras] {10.1093/mnras/stac733},
  \href {https://ui.adsabs.harvard.edu/abs/2022MNRAS.512.4798C} {512, 4798}

\bibitem[\protect\citeauthoryear{{Cullinane} et~al.,}{{Cullinane}
  et~al.}{2023}]{Cullinane2023}
{Cullinane} L.~R.,  et~al., 2023, \mn@doi [\apj] {10.3847/1538-4357/ad003b},
  \href {https://scixplorer.org/abs/2023ApJ...958..157C} {958, 157}

\bibitem[\protect\citeauthoryear{{Danhaive} et~al.,}{{Danhaive}
  et~al.}{2025}]{Danhaive2025}
{Danhaive} A.~L.,  et~al., 2025, \mn@doi [\mnras] {10.1093/mnras/staf1540},
  \href {https://scixplorer.org/abs/2025MNRAS.543.3249D} {543, 3249}

\bibitem[\protect\citeauthoryear{{Dekel}, {Ginzburg}, {Jiang}, {Freundlich},
  {Lapiner}, {Ceverino}  \& {Primack}}{{Dekel} et~al.}{2020}]{Dekel2020}
{Dekel} A.,  {Ginzburg} O.,  {Jiang} F.,  {Freundlich} J.,  {Lapiner} S.,
  {Ceverino} D.,   {Primack} J.,  2020, \mn@doi [\mnras]
  {10.1093/mnras/staa470}, \href
  {https://scixplorer.org/abs/2020MNRAS.493.4126D} {493, 4126}

\bibitem[\protect\citeauthoryear{{Di Matteo}, {Haywood}, {Lehnert}, {Katz},
  {Khoperskov}, {Snaith}, {G{\'o}mez}  \& {Robichon}}{{Di Matteo}
  et~al.}{2019}]{DiMatteo2019}
{Di Matteo} P.,  {Haywood} M.,  {Lehnert} M.~D.,  {Katz} D.,  {Khoperskov} S.,
  {Snaith} O.~N.,  {G{\'o}mez} A.,   {Robichon} N.,  2019, \mn@doi [\aap]
  {10.1051/0004-6361/201834929}, \href
  {https://scixplorer.org/abs/2019A&A...632A...4D} {632, A4}

\bibitem[\protect\citeauthoryear{{Dillamore}, {Belokurov}, {Kravtsov}  \&
  {Font}}{{Dillamore} et~al.}{2024}]{Dillamore2024}
{Dillamore} A.~M.,  {Belokurov} V.,  {Kravtsov} A.,   {Font} A.~S.,  2024,
  \mn@doi [\mnras] {10.1093/mnras/stad3369}, \href
  {https://scixplorer.org/abs/2024MNRAS.527.7070D} {527, 7070}

\bibitem[\protect\citeauthoryear{{El-Badry} et~al.,}{{El-Badry}
  et~al.}{2018}]{ElBadry2018}
{El-Badry} K.,  et~al., 2018, \mn@doi [\mnras] {10.1093/mnras/stx2482}, \href
  {https://scixplorer.org/abs/2018MNRAS.473.1930E} {473, 1930}

\bibitem[\protect\citeauthoryear{{Fitts} et~al.,}{{Fitts}
  et~al.}{2018}]{Fitts2018}
{Fitts} A.,  et~al., 2018, \mn@doi [\mnras] {10.1093/mnras/sty1488}, \href
  {https://scixplorer.org/abs/2018MNRAS.479..319F} {479, 319}

\bibitem[\protect\citeauthoryear{{Foreman-Mackey}}{{Foreman-Mackey}}{2016}]{ForemanMackey2016}
{Foreman-Mackey} D.,  2016, \mn@doi [The Journal of Open Source Software]
  {10.21105/joss.00024}, \href
  {https://ui.adsabs.harvard.edu/abs/2016JOSS....1...24F} {1, 24}

\bibitem[\protect\citeauthoryear{{Freeman}, {Illingworth}  \&
  {Oemler}}{{Freeman} et~al.}{1983}]{Freeman1983}
{Freeman} K.~C.,  {Illingworth} G.,   {Oemler} Jr. A.,  1983, \mn@doi [\apj]
  {10.1086/161316}, \href {https://scixplorer.org/abs/1983ApJ...272..488F}
  {272, 488}

\bibitem[\protect\citeauthoryear{{Gaia Collaboration} et~al.,}{{Gaia
  Collaboration} et~al.}{2023}]{GaiaCollaboration2023}
{Gaia Collaboration} et~al., 2023, \mn@doi [\aap]
  {10.1051/0004-6361/202243940}, \href
  {https://ui.adsabs.harvard.edu/abs/2023A&A...674A...1G} {674, A1}

\bibitem[\protect\citeauthoryear{{Gallart}, {Stetson}, {Hardy}, {Pont}  \&
  {Zinn}}{{Gallart} et~al.}{2004}]{Gallart2004}
{Gallart} C.,  {Stetson} P.~B.,  {Hardy} E.,  {Pont} F.,   {Zinn} R.,  2004,
  \mn@doi [\apjl] {10.1086/425866}, \href
  {https://scixplorer.org/abs/2004ApJ...614L.109G} {614, L109}

\bibitem[\protect\citeauthoryear{{Gallart}, {Bernard}, {Brook}, {Ruiz-Lara},
  {Cassisi}, {Hill}  \& {Monelli}}{{Gallart} et~al.}{2019}]{Gallart2019}
{Gallart} C.,  {Bernard} E.~J.,  {Brook} C.~B.,  {Ruiz-Lara} T.,  {Cassisi} S.,
   {Hill} V.,   {Monelli} M.,  2019, \mn@doi [Nature Astronomy]
  {10.1038/s41550-019-0829-5}, \href
  {https://scixplorer.org/abs/2019NatAs...3..932G} {3, 932}

\bibitem[\protect\citeauthoryear{{Geda}, {Cruz}, {Wright}, {Greene}, {Brooks},
  {Quinn}, {Wadsley}  \& {Keller}}{{Geda} et~al.}{2026}]{Geda2026}
{Geda} R.,  {Cruz} A.,  {Wright} A.~C.,  {Greene} J.~E.,  {Brooks} A.,  {Quinn}
  T.,  {Wadsley} J.,   {Keller} B.,  2026, \mn@doi [\apj]
  {10.3847/1538-4357/ae64fd}, \href
  {https://ui.adsabs.harvard.edu/abs/2026ApJ..1004..110G} {1004, 110}

\bibitem[\protect\citeauthoryear{{Gilbert} et~al.,}{{Gilbert}
  et~al.}{2022}]{Gilbert2022}
{Gilbert} K.~M.,  et~al., 2022, \mn@doi [\apj] {10.3847/1538-4357/ac3480},
  \href {https://scixplorer.org/abs/2022ApJ...924..116G} {924, 116}

\bibitem[\protect\citeauthoryear{{Ginsburg} et~al.,}{{Ginsburg}
  et~al.}{2019}]{Ginsburg2019}
{Ginsburg} A.,  et~al., 2019, \mn@doi [\aj] {10.3847/1538-3881/aafc33}, \href
  {https://ui.adsabs.harvard.edu/abs/2019AJ....157...98G} {157, 98}

\bibitem[\protect\citeauthoryear{{Girard} et~al.,}{{Girard}
  et~al.}{2021}]{Girard2021}
{Girard} M.,  et~al., 2021, \mn@doi [\apj] {10.3847/1538-4357/abd5b9}, \href
  {https://scixplorer.org/abs/2021ApJ...909...12G} {909, 12}

\bibitem[\protect\citeauthoryear{{G{\'o}rski}, {Hivon}, {Banday}, {Wandelt},
  {Hansen}, {Reinecke}  \& {Bartelmann}}{{G{\'o}rski}
  et~al.}{2005}]{Gorski2005}
{G{\'o}rski} K.~M.,  {Hivon} E.,  {Banday} A.~J.,  {Wandelt} B.~D.,  {Hansen}
  F.~K.,  {Reinecke} M.,   {Bartelmann} M.,  2005, \mn@doi [\apj]
  {10.1086/427976}, \href
  {https://ui.adsabs.harvard.edu/abs/2005ApJ...622..759G} {622, 759}

\bibitem[\protect\citeauthoryear{{Grady}, {Belokurov}  \& {Evans}}{{Grady}
  et~al.}{2021}]{Grady2021}
{Grady} J.,  {Belokurov} V.,   {Evans} N.~W.,  2021, \mn@doi [\apj]
  {10.3847/1538-4357/abd4e4}, \href
  {https://ui.adsabs.harvard.edu/abs/2021ApJ...909..150G} {909, 150}

\bibitem[\protect\citeauthoryear{{Grand} et~al.,}{{Grand}
  et~al.}{2017}]{Grand2017}
{Grand} R.~J.~J.,  et~al., 2017, \mn@doi [\mnras] {10.1093/mnras/stx071}, \href
  {https://ui.adsabs.harvard.edu/abs/2017MNRAS.467..179G} {467, 179}

\bibitem[\protect\citeauthoryear{{Grand}, {Fragkoudi}, {G{\'o}mez}, {Jenkins},
  {Marinacci}, {Pakmor}  \& {Springel}}{{Grand} et~al.}{2024}]{Grand2024}
{Grand} R. J.~J.,  {Fragkoudi} F.,  {G{\'o}mez} F.~A.,  {Jenkins} A.,
  {Marinacci} F.,  {Pakmor} R.,   {Springel} V.,  2024, \mn@doi [\mnras]
  {10.1093/mnras/stae1598}, \href
  {https://ui.adsabs.harvard.edu/abs/2024MNRAS.532.1814G} {532, 1814}

\bibitem[\protect\citeauthoryear{{Gratton}, {Bragaglia}, {Clementini},
  {Carretta}, {Di Fabrizio}, {Maio}  \& {Taribello}}{{Gratton}
  et~al.}{2004}]{Gratton2004}
{Gratton} R.~G.,  {Bragaglia} A.,  {Clementini} G.,  {Carretta} E.,  {Di
  Fabrizio} L.,  {Maio} M.,   {Taribello} E.,  2004, \mn@doi [\aap]
  {10.1051/0004-6361:20035840}, \href
  {https://ui.adsabs.harvard.edu/abs/2004A&A...421..937G} {421, 937}

\bibitem[\protect\citeauthoryear{{Green}}{{Green}}{2018}]{Green2018}
{Green} G.~M.,  2018, \mn@doi [The Journal of Open Source Software]
  {10.21105/joss.00695}, \href
  {https://ui.adsabs.harvard.edu/abs/2018JOSS....3..695G} {3, 695}

\bibitem[\protect\citeauthoryear{{Gunn} et~al.,}{{Gunn}
  et~al.}{2006}]{Gunn2006}
{Gunn} J.~E.,  et~al., 2006, \mn@doi [\aj] {10.1086/500975}, \href
  {https://ui.adsabs.harvard.edu/abs/2006AJ....131.2332G} {131, 2332}

\bibitem[\protect\citeauthoryear{{Hafen} et~al.,}{{Hafen}
  et~al.}{2022}]{Hafen2022}
{Hafen} Z.,  et~al., 2022, \mn@doi [\mnras] {10.1093/mnras/stac1603}, \href
  {https://scixplorer.org/abs/2022MNRAS.514.5056H} {514, 5056}

\bibitem[\protect\citeauthoryear{{Harris} et~al.,}{{Harris}
  et~al.}{2020}]{Harris2020}
{Harris} C.~R.,  et~al., 2020, \mn@doi [\nat] {10.1038/s41586-020-2649-2},
  \href {https://ui.adsabs.harvard.edu/abs/2020Natur.585..357H} {585, 357}

\bibitem[\protect\citeauthoryear{{Hopkins}}{{Hopkins}}{2015}]{Hopkins2015}
{Hopkins} P.~F.,  2015, \mn@doi [\mnras] {10.1093/mnras/stv195}, \href
  {https://ui.adsabs.harvard.edu/abs/2015MNRAS.450...53H} {450, 53}

\bibitem[\protect\citeauthoryear{{Hopkins} et~al.,}{{Hopkins}
  et~al.}{2018}]{Hopkins2018}
{Hopkins} P.~F.,  et~al., 2018, \mn@doi [\mnras] {10.1093/mnras/sty1690}, \href
  {https://ui.adsabs.harvard.edu/abs/2018MNRAS.480..800H} {480, 800}

\bibitem[\protect\citeauthoryear{{Hopkins} et~al.,}{{Hopkins}
  et~al.}{2023}]{Hopkins2023}
{Hopkins} P.~F.,  et~al., 2023, \mn@doi [\mnras] {10.1093/mnras/stad1902},
  \href {https://scixplorer.org/abs/2023MNRAS.525.2241H} {525, 2241}

\bibitem[\protect\citeauthoryear{{Hunter}}{{Hunter}}{2007}]{Hunter2007}
{Hunter} J.~D.,  2007, \mn@doi [Computing in Science and Engineering]
  {10.1109/MCSE.2007.55}, \href
  {https://ui.adsabs.harvard.edu/abs/2007CSE.....9...90H} {9, 90}

\bibitem[\protect\citeauthoryear{{Kinman}, {Stryker}, {Hesser}, {Graham},
  {Walker}, {Hazen}  \& {Nemec}}{{Kinman} et~al.}{1991}]{Kinman1991}
{Kinman} T.~D.,  {Stryker} L.~L.,  {Hesser} J.~E.,  {Graham} J.~A.,  {Walker}
  A.~R.,  {Hazen} M.~L.,   {Nemec} J.~M.,  1991, \mn@doi [\pasp]
  {10.1086/132947}, \href {https://scixplorer.org/abs/1991PASP..103.1279K}
  {103, 1279}

\bibitem[\protect\citeauthoryear{{Kohandel}, {Pallottini}, {Ferrara},
  {Zanella}, {Rizzo}  \& {Carniani}}{{Kohandel} et~al.}{2024}]{Kohandel2024}
{Kohandel} M.,  {Pallottini} A.,  {Ferrara} A.,  {Zanella} A.,  {Rizzo} F.,
  {Carniani} S.,  2024, \mn@doi [\aap] {10.1051/0004-6361/202348209}, \href
  {https://scixplorer.org/abs/2024A&A...685A..72K} {685, A72}

\bibitem[\protect\citeauthoryear{{Kohandel}, {Pallottini}  \&
  {Ferrara}}{{Kohandel} et~al.}{2025}]{Kohandel2025}
{Kohandel} M.,  {Pallottini} A.,   {Ferrara} A.,  2025, \mn@doi [\aap]
  {10.1051/0004-6361/202555499}, \href
  {https://scixplorer.org/abs/2025A&A...704A..39K} {704, A39}

\bibitem[\protect\citeauthoryear{{Kollmeier} et~al.,}{{Kollmeier}
  et~al.}{2026}]{Kollmeier2026}
{Kollmeier} J.~A.,  et~al., 2026, \mn@doi [\aj] {10.3847/1538-3881/ae0576},
  \href {https://ui.adsabs.harvard.edu/abs/2026AJ....171...52K} {171, 52}

\bibitem[\protect\citeauthoryear{{Koposov} \& {Bartunov}}{{Koposov} \&
  {Bartunov}}{2006}]{Koposov2006}
{Koposov} S.,  {Bartunov} O.,  2006, in {Gabriel} C.,  {Arviset} C.,  {Ponz}
  D.,   {Enrique} S.,  eds,  Astronomical Society of the Pacific Conference
  Series Vol. 351, Astronomical Data Analysis Software and Systems XV. p.~735

\bibitem[\protect\citeauthoryear{{Kroupa}}{{Kroupa}}{2001}]{Kroupa2001}
{Kroupa} P.,  2001, \mn@doi [\mnras] {10.1046/j.1365-8711.2001.04022.x}, \href
  {https://ui.adsabs.harvard.edu/abs/2001MNRAS.322..231K} {322, 231}

\bibitem[\protect\citeauthoryear{{Kurbatov} et~al.,}{{Kurbatov}
  et~al.}{2024}]{Kurbatov2024}
{Kurbatov} E.~P.,  et~al., 2024, \mn@doi [arXiv e-prints]
  {10.48550/arXiv.2410.22250}, \href
  {https://scixplorer.org/abs/2024arXiv241022250K} {p. arXiv:2410.22250}

\bibitem[\protect\citeauthoryear{{Leaman}, {Cole}, {Venn}, {Tolstoy}, {Irwin},
  {Szeifert}, {Skillman}  \& {McConnachie}}{{Leaman} et~al.}{2009}]{Leaman2009}
{Leaman} R.,  {Cole} A.~A.,  {Venn} K.~A.,  {Tolstoy} E.,  {Irwin} M.~J.,
  {Szeifert} T.,  {Skillman} E.~D.,   {McConnachie} A.~W.,  2009, \mn@doi
  [\apj] {10.1088/0004-637X/699/1/1}, \href
  {https://scixplorer.org/abs/2009ApJ...699....1L} {699, 1}

\bibitem[\protect\citeauthoryear{{Mackey}, {Koposov}, {Erkal}, {Belokurov}, {Da
  Costa}  \& {G{\'o}mez}}{{Mackey} et~al.}{2016}]{Mackey2016}
{Mackey} A.~D.,  {Koposov} S.~E.,  {Erkal} D.,  {Belokurov} V.,  {Da Costa}
  G.~S.,   {G{\'o}mez} F.~A.,  2016, \mn@doi [\mnras] {10.1093/mnras/stw497},
  \href {https://ui.adsabs.harvard.edu/abs/2016MNRAS.459..239M} {459, 239}

\bibitem[\protect\citeauthoryear{{Mackey}, {Koposov}, {Da Costa}, {Belokurov},
  {Erkal}  \& {Kuzma}}{{Mackey} et~al.}{2018}]{Mackey2018}
{Mackey} D.,  {Koposov} S.,  {Da Costa} G.,  {Belokurov} V.,  {Erkal} D.,
  {Kuzma} P.,  2018, \mn@doi [\apjl] {10.3847/2041-8213/aac175}, \href
  {https://ui.adsabs.harvard.edu/abs/2018ApJ...858L..21M} {858, L21}

\bibitem[\protect\citeauthoryear{{Majewski}, {Nidever}, {Mu{\~n}oz},
  {Patterson}, {Kunkel}  \& {Carlin}}{{Majewski} et~al.}{2009}]{Majewski2009}
{Majewski} S.~R.,  {Nidever} D.~L.,  {Mu{\~n}oz} R.~R.,  {Patterson} R.~J.,
  {Kunkel} W.~E.,   {Carlin} J.~L.,  2009, in {Van Loon} J.~T.,  {Oliveira}
  J.~M.,  eds,  IAU Symposium Vol. 256, The Magellanic System: Stars, Gas, and
  Galaxies. pp 51--56, \mn@doi{10.1017/S1743921308028251}

\bibitem[\protect\citeauthoryear{{Majewski} et~al.,}{{Majewski}
  et~al.}{2017}]{Majewski2017}
{Majewski} S.~R.,  et~al., 2017, \mn@doi [\aj] {10.3847/1538-3881/aa784d},
  \href {https://ui.adsabs.harvard.edu/abs/2017AJ....154...94M} {154, 94}

\bibitem[\protect\citeauthoryear{{Marigo} et~al.,}{{Marigo}
  et~al.}{2017}]{Marigo2017}
{Marigo} P.,  et~al., 2017, \mn@doi [\apj] {10.3847/1538-4357/835/1/77}, \href
  {https://ui.adsabs.harvard.edu/abs/2017ApJ...835...77M} {835, 77}

\bibitem[\protect\citeauthoryear{{McCarthy}, {Font}, {Crain}, {Deason},
  {Schaye}  \& {Theuns}}{{McCarthy} et~al.}{2012}]{McCarthy2012}
{McCarthy} I.~G.,  {Font} A.~S.,  {Crain} R.~A.,  {Deason} A.~J.,  {Schaye} J.,
    {Theuns} T.,  2012, \mn@doi [\mnras] {10.1111/j.1365-2966.2011.20189.x},
  \href {https://scixplorer.org/abs/2012MNRAS.420.2245M} {420, 2245}

\bibitem[\protect\citeauthoryear{{McKinney}}{{McKinney}}{2010}]{McKinney2010}
{McKinney} W.,  2010, in Proceedings of the 9th Python in Science Conference.
  pp 56--61, \mn@doi{10.25080/Majora-92bf1922-00a}

\bibitem[\protect\citeauthoryear{{Minniti}, {Borissova}, {Rejkuba}, {Alves},
  {Cook}  \& {Freeman}}{{Minniti} et~al.}{2003}]{Minniti2003}
{Minniti} D.,  {Borissova} J.,  {Rejkuba} M.,  {Alves} D.~R.,  {Cook} K.~H.,
  {Freeman} K.~C.,  2003, \mn@doi [Science] {10.1126/science.1088529}, \href
  {https://scixplorer.org/abs/2003Sci...301.1508M} {301, 1508}

\bibitem[\protect\citeauthoryear{{Mucciarelli}, {Massari}, {Minelli}, {Romano},
  {Bellazzini}, {Ferraro}, {Matteucci}  \& {Origlia}}{{Mucciarelli}
  et~al.}{2021}]{Mucciarelli2021}
{Mucciarelli} A.,  {Massari} D.,  {Minelli} A.,  {Romano} D.,  {Bellazzini} M.,
   {Ferraro} F.~R.,  {Matteucci} F.,   {Origlia} L.,  2021, \mn@doi [Nature
  Astronomy] {10.1038/s41550-021-01493-y}, \href
  {https://scixplorer.org/abs/2021NatAs...5.1247M} {5, 1247}

\bibitem[\protect\citeauthoryear{{Nadler}, {Kong}, {Yang}  \& {Yu}}{{Nadler}
  et~al.}{2025a}]{Nadler2025data}
{Nadler} E.,  {Kong} D.,  {Yang} D.,   {Yu} H.-B.,  2025a, {Data Release: SIDM
  Concerto: Compilation and Data Release of Self-interacting Dark Matter
  Zoom-in Simulations}, \mn@doi{10.5281/zenodo.14933624}, \url
  {https://doi.org/10.5281/zenodo.14933624}

\bibitem[\protect\citeauthoryear{{Nadler}, {Kong}, {Yang}  \& {Yu}}{{Nadler}
  et~al.}{2025b}]{Nadler2025}
{Nadler} E.~O.,  {Kong} D.,  {Yang} D.,   {Yu} H.-B.,  2025b, \mn@doi [\apj]
  {10.3847/1538-4357/adf553}, \href
  {https://ui.adsabs.harvard.edu/abs/2025ApJ...991...69N} {991, 69}

\bibitem[\protect\citeauthoryear{{Navarrete} et~al.,}{{Navarrete}
  et~al.}{2019}]{Navarrete2019}
{Navarrete} C.,  et~al., 2019, \mn@doi [\mnras] {10.1093/mnras/sty3347}, \href
  {https://scixplorer.org/abs/2019MNRAS.483.4160N} {483, 4160}

\bibitem[\protect\citeauthoryear{{Navarrete}, {Aguado}, {Belokurov}, {Erkal},
  {Deason}, {Cullinane}  \& {Carballo-Bello}}{{Navarrete}
  et~al.}{2023}]{Navarrete2023}
{Navarrete} C.,  {Aguado} D.~S.,  {Belokurov} V.,  {Erkal} D.,  {Deason} A.,
  {Cullinane} L.,   {Carballo-Bello} J.,  2023, \mn@doi [\mnras]
  {10.1093/mnras/stad1698}, \href
  {https://scixplorer.org/abs/2023MNRAS.523.4720N} {523, 4720}

\bibitem[\protect\citeauthoryear{{Nelson} et~al.,}{{Nelson}
  et~al.}{2019a}]{Nelson2019release}
{Nelson} D.,  et~al., 2019a, \mn@doi [Computational Astrophysics and Cosmology]
  {10.1186/s40668-019-0028-x}, \href
  {https://ui.adsabs.harvard.edu/abs/2019ComAC...6....2N} {6, 2}

\bibitem[\protect\citeauthoryear{{Nelson} et~al.,}{{Nelson}
  et~al.}{2019b}]{Nelson2019TNG50}
{Nelson} D.,  et~al., 2019b, \mn@doi [\mnras] {10.1093/mnras/stz2306}, \href
  {https://ui.adsabs.harvard.edu/abs/2019MNRAS.490.3234N} {490, 3234}

\bibitem[\protect\citeauthoryear{{Nidever} et~al.,}{{Nidever}
  et~al.}{2019}]{Nidever2019}
{Nidever} D.~L.,  et~al., 2019, \mn@doi [\apj] {10.3847/1538-4357/aafaf7},
  \href {https://scixplorer.org/abs/2019ApJ...874..118N} {874, 118}

\bibitem[\protect\citeauthoryear{{Nidever} et~al.,}{{Nidever}
  et~al.}{2026}]{Nidever2026}
{Nidever} D.~L.,  et~al., 2026, \mn@doi [\aj] {10.3847/1538-3881/ae7c76}, \href
  {https://ui.adsabs.harvard.edu/abs/2026AJ....172..130N} {172, 130}

\bibitem[\protect\citeauthoryear{{Ochsenbein}, {Bauer}  \&
  {Marcout}}{{Ochsenbein} et~al.}{2000}]{Ochsenbein2000}
{Ochsenbein} F.,  {Bauer} P.,   {Marcout} J.,  2000, \mn@doi [\aaps]
  {10.1051/aas:2000169}, \href
  {https://ui.adsabs.harvard.edu/abs/2000A&AS..143...23O} {143, 23}

\bibitem[\protect\citeauthoryear{{Olsen}, {Hodge}, {Mateo}, {Olszewski},
  {Schommer}, {Suntzeff}  \& {Walker}}{{Olsen} et~al.}{1998}]{Olsen1998}
{Olsen} K.~A.~G.,  {Hodge} P.~W.,  {Mateo} M.,  {Olszewski} E.~W.,  {Schommer}
  R.~A.,  {Suntzeff} N.~B.,   {Walker} A.~R.,  1998, \mn@doi [\mnras]
  {10.1046/j.1365-8711.1998.01860.x}, \href
  {https://scixplorer.org/abs/1998MNRAS.300..665O} {300, 665}

\bibitem[\protect\citeauthoryear{{Pejcha} \& {Stanek}}{{Pejcha} \&
  {Stanek}}{2009}]{Pejcha2009}
{Pejcha} O.,  {Stanek} K.~Z.,  2009, \mn@doi [\apj]
  {10.1088/0004-637X/704/2/1730}, \href
  {https://scixplorer.org/abs/2009ApJ...704.1730P} {704, 1730}

\bibitem[\protect\citeauthoryear{{Peter}, {Rocha}, {Bullock}  \&
  {Kaplinghat}}{{Peter} et~al.}{2013}]{Peter2013}
{Peter} A. H.~G.,  {Rocha} M.,  {Bullock} J.~S.,   {Kaplinghat} M.,  2013,
  \mn@doi [\mnras] {10.1093/mnras/sts535}, \href
  {https://ui.adsabs.harvard.edu/abs/2013MNRAS.430..105P} {430, 105}

\bibitem[\protect\citeauthoryear{{Phan}, {Pradhan}  \& {Jankowiak}}{{Phan}
  et~al.}{2019}]{Phan2019}
{Phan} D.,  {Pradhan} N.,   {Jankowiak} M.,  2019, \mn@doi [arXiv e-prints]
  {10.48550/arXiv.1912.11554}, \href
  {https://ui.adsabs.harvard.edu/abs/2019arXiv191211554P} {p. arXiv:1912.11554}

\bibitem[\protect\citeauthoryear{{Pillepich} et~al.,}{{Pillepich}
  et~al.}{2019}]{Pillepich2019}
{Pillepich} A.,  et~al., 2019, \mn@doi [\mnras] {10.1093/mnras/stz2338}, \href
  {https://ui.adsabs.harvard.edu/abs/2019MNRAS.490.3196P} {490, 3196}

\bibitem[\protect\citeauthoryear{{Povick} et~al.,}{{Povick}
  et~al.}{2024}]{Povick2024}
{Povick} J.~T.,  et~al., 2024, \mn@doi [\mnras] {10.1093/mnras/stae1949}, \href
  {https://ui.adsabs.harvard.edu/abs/2024MNRAS.533.3685P} {533, 3685}

\bibitem[\protect\citeauthoryear{{Price-Whelan}}{{Price-Whelan}}{2017}]{PriceWhelan2017}
{Price-Whelan} A.~M.,  2017, \mn@doi [The Journal of Open Source Software]
  {10.21105/joss.00388}, \href
  {https://ui.adsabs.harvard.edu/abs/2017JOSS....2..388P} {2, 388}

\bibitem[\protect\citeauthoryear{{Rix} et~al.,}{{Rix} et~al.}{2022}]{Rix2022}
{Rix} H.-W.,  et~al., 2022, \mn@doi [\apj] {10.3847/1538-4357/ac9e01}, \href
  {https://scixplorer.org/abs/2022ApJ...941...45R} {941, 45}

\bibitem[\protect\citeauthoryear{{Rizzo}, {Vegetti}, {Powell}, {Fraternali},
  {McKean}, {Stacey}  \& {White}}{{Rizzo} et~al.}{2020}]{Rizzo2020}
{Rizzo} F.,  {Vegetti} S.,  {Powell} D.,  {Fraternali} F.,  {McKean} J.~P.,
  {Stacey} H.~R.,   {White} S.~D.~M.,  2020, \mn@doi [\nat]
  {10.1038/s41586-020-2572-6}, \href
  {https://scixplorer.org/abs/2020Natur.584..201R} {584, 201}

\bibitem[\protect\citeauthoryear{{Rizzo} et~al.,}{{Rizzo}
  et~al.}{2024}]{Rizzo2024}
{Rizzo} F.,  et~al., 2024, \mn@doi [\aap] {10.1051/0004-6361/202450455}, \href
  {https://scixplorer.org/abs/2024A&A...689A.273R} {689, A273}

\bibitem[\protect\citeauthoryear{{Rowland} et~al.,}{{Rowland}
  et~al.}{2024}]{Rowland2024}
{Rowland} L.~E.,  et~al., 2024, \mn@doi [\mnras] {10.1093/mnras/stae2217},
  \href {https://scixplorer.org/abs/2024MNRAS.535.2068R} {535, 2068}

\bibitem[\protect\citeauthoryear{{Ruan} et~al.,}{{Ruan}
  et~al.}{2026}]{Ruan2026}
{Ruan} D.,  et~al., 2026, \mn@doi [arXiv e-prints] {10.48550/arXiv.2605.06893},
  \href {https://ui.adsabs.harvard.edu/abs/2026arXiv260506893R} {p.
  arXiv:2605.06893}

\bibitem[\protect\citeauthoryear{{Saha} et~al.,}{{Saha}
  et~al.}{2010}]{Saha2010}
{Saha} A.,  et~al., 2010, \mn@doi [\aj] {10.1088/0004-6256/140/6/1719}, \href
  {https://scixplorer.org/abs/2010AJ....140.1719S} {140, 1719}

\bibitem[\protect\citeauthoryear{{Sales}, {Navarro}, {Theuns}, {Schaye},
  {White}, {Frenk}, {Crain}  \& {Dalla Vecchia}}{{Sales}
  et~al.}{2012}]{Sales2012}
{Sales} L.~V.,  {Navarro} J.~F.,  {Theuns} T.,  {Schaye} J.,  {White} S. D.~M.,
   {Frenk} C.~S.,  {Crain} R.~A.,   {Dalla Vecchia} C.,  2012, \mn@doi [\mnras]
  {10.1111/j.1365-2966.2012.20975.x}, \href
  {https://scixplorer.org/abs/2012MNRAS.423.1544S} {423, 1544}

\bibitem[\protect\citeauthoryear{{Sameie}, {Creasey}, {Yu}, {Sales},
  {Vogelsberger}  \& {Zavala}}{{Sameie} et~al.}{2018}]{Sameie2018}
{Sameie} O.,  {Creasey} P.,  {Yu} H.-B.,  {Sales} L.~V.,  {Vogelsberger} M.,
  {Zavala} J.,  2018, \mn@doi [\mnras] {10.1093/mnras/sty1516}, \href
  {https://scixplorer.org/abs/2018MNRAS.479..359S} {479, 359}

\bibitem[\protect\citeauthoryear{{Schommer}, {Suntzeff}, {Olszewski}  \&
  {Harris}}{{Schommer} et~al.}{1992}]{Schommer1992}
{Schommer} R.~A.,  {Suntzeff} N.~B.,  {Olszewski} E.~W.,   {Harris} H.~C.,
  1992, \mn@doi [\aj] {10.1086/116074}, \href
  {https://scixplorer.org/abs/1992AJ....103..447S} {103, 447}

\bibitem[\protect\citeauthoryear{{Semenov}}{{Semenov}}{2026}]{Semenov2026}
{Semenov} V.~A.,  2026, \mn@doi [\apj] {10.3847/1538-4357/ae83aa}, \href
  {https://scixplorer.org/abs/2026ApJ..1010..123S} {1010, 123}

\bibitem[\protect\citeauthoryear{{Semenov}, {Conroy}, {Chandra}, {Hernquist}
  \& {Nelson}}{{Semenov} et~al.}{2024}]{Semenov2024b}
{Semenov} V.~A.,  {Conroy} C.,  {Chandra} V.,  {Hernquist} L.,   {Nelson} D.,
  2024, \mn@doi [\apj] {10.3847/1538-4357/ad57ba}, \href
  {https://scixplorer.org/abs/2024ApJ...972...73S} {972, 73}

\bibitem[\protect\citeauthoryear{{Semenov}, {Conroy}, {Smith}, {Puchwein}  \&
  {Hernquist}}{{Semenov} et~al.}{2025}]{Semenov2025a}
{Semenov} V.~A.,  {Conroy} C.,  {Smith} A.,  {Puchwein} E.,   {Hernquist} L.,
  2025, \mn@doi [\apj] {10.3847/1538-4357/addf48}, \href
  {https://scixplorer.org/abs/2025ApJ...990....7S} {990, 7}

\bibitem[\protect\citeauthoryear{{Simons} et~al.,}{{Simons}
  et~al.}{2016}]{Simons2016}
{Simons} R.~C.,  et~al., 2016, \mn@doi [\apj] {10.3847/0004-637X/830/1/14},
  \href {https://scixplorer.org/abs/2016ApJ...830...14S} {830, 14}

\bibitem[\protect\citeauthoryear{{Sizemore}, {Llanes}, {Kounkel}, {Hutchinson},
  {Stassun}  \& {Chandra}}{{Sizemore} et~al.}{2024}]{Sizemore2024}
{Sizemore} L.,  {Llanes} D.,  {Kounkel} M.,  {Hutchinson} B.,  {Stassun} K.~G.,
    {Chandra} V.,  2024, \mn@doi [\aj] {10.3847/1538-3881/ad291d}, \href
  {https://ui.adsabs.harvard.edu/abs/2024AJ....167..173S} {167, 173}

\bibitem[\protect\citeauthoryear{{Skowron} et~al.,}{{Skowron}
  et~al.}{2016}]{Skowron2016}
{Skowron} D.~M.,  et~al., 2016, Acta Astron., \href
  {https://ui.adsabs.harvard.edu/abs/2016AcA....66..269S} {66, 269}

\bibitem[\protect\citeauthoryear{{Smee} et~al.,}{{Smee}
  et~al.}{2013}]{Smee2013}
{Smee} S.~A.,  et~al., 2013, \mn@doi [\aj] {10.1088/0004-6256/146/2/32}, \href
  {https://ui.adsabs.harvard.edu/abs/2013AJ....146...32S} {146, 32}

\bibitem[\protect\citeauthoryear{{Soszy{\'n}ski} et~al.,}{{Soszy{\'n}ski}
  et~al.}{2016}]{Soszynski2016}
{Soszy{\'n}ski} I.,  et~al., 2016, \mn@doi [\actaa]
  {10.48550/arXiv.1606.02727}, \href
  {https://ui.adsabs.harvard.edu/abs/2016AcA....66..131S} {66, 131}

\bibitem[\protect\citeauthoryear{{Soszy{\'n}ski} et~al.,}{{Soszy{\'n}ski}
  et~al.}{2019}]{Soszynski2019}
{Soszy{\'n}ski} I.,  et~al., 2019, \mn@doi [\actaa]
  {10.32023/0001-5237/69.2.1}, \href
  {https://ui.adsabs.harvard.edu/abs/2019AcA....69...87S} {69, 87}

\bibitem[\protect\citeauthoryear{{Springel} \& {Hernquist}}{{Springel} \&
  {Hernquist}}{2003}]{Springel2003}
{Springel} V.,  {Hernquist} L.,  2003, \mn@doi [\mnras]
  {10.1046/j.1365-8711.2003.06206.x}, \href
  {https://scixplorer.org/abs/2003MNRAS.339..289S} {339, 289}

\bibitem[\protect\citeauthoryear{{Stern} et~al.,}{{Stern}
  et~al.}{2021}]{Stern2021}
{Stern} J.,  et~al., 2021, \mn@doi [\apj] {10.3847/1538-4357/abd776}, \href
  {https://scixplorer.org/abs/2021ApJ...911...88S} {911, 88}

\bibitem[\protect\citeauthoryear{{Swan}, {Cole}, {Tolstoy}  \& {Irwin}}{{Swan}
  et~al.}{2016}]{Swan2016}
{Swan} J.,  {Cole} A.~A.,  {Tolstoy} E.,   {Irwin} M.~J.,  2016, \mn@doi
  [\mnras] {10.1093/mnras/stv2774}, \href
  {https://scixplorer.org/abs/2016MNRAS.456.4315S} {456, 4315}

\bibitem[\protect\citeauthoryear{{Udalski}, {Szyma{\'n}ski}  \&
  {Szyma{\'n}ski}}{{Udalski} et~al.}{2015}]{Udalski2015}
{Udalski} A.,  {Szyma{\'n}ski} M.~K.,   {Szyma{\'n}ski} G.,  2015, \mn@doi
  [\actaa] {10.48550/arXiv.1504.05966}, \href
  {https://ui.adsabs.harvard.edu/abs/2015AcA....65....1U} {65, 1}

\bibitem[\protect\citeauthoryear{{Vargya}, {Sanderson}, {Sameie},
  {Boylan-Kolchin}, {Hopkins}, {Wetzel}  \& {Graus}}{{Vargya}
  et~al.}{2022}]{Vargya2022}
{Vargya} D.,  {Sanderson} R.,  {Sameie} O.,  {Boylan-Kolchin} M.,  {Hopkins}
  P.~F.,  {Wetzel} A.,   {Graus} A.,  2022, \mn@doi [\mnras]
  {10.1093/mnras/stac2069}, \href
  {https://scixplorer.org/abs/2022MNRAS.516.2389V} {516, 2389}

\bibitem[\protect\citeauthoryear{{Vasiliev}}{{Vasiliev}}{2018}]{Vasiliev2018}
{Vasiliev} E.,  2018, \mn@doi [\mnras] {10.1093/mnrasl/sly168}, \href
  {https://scixplorer.org/abs/2018MNRAS.481L.100V} {481, L100}

\bibitem[\protect\citeauthoryear{{Vasiliev}}{{Vasiliev}}{2019}]{Vasiliev2019}
{Vasiliev} E.,  2019, \mn@doi [\mnras] {10.1093/mnras/sty2672}, \href
  {https://ui.adsabs.harvard.edu/abs/2019MNRAS.482.1525V} {482, 1525}

\bibitem[\protect\citeauthoryear{{Vasiliev} \& {Belokurov}}{{Vasiliev} \&
  {Belokurov}}{2020}]{Vasiliev2020}
{Vasiliev} E.,  {Belokurov} V.,  2020, \mn@doi [\mnras]
  {10.1093/mnras/staa2114}, \href
  {https://scixplorer.org/abs/2020MNRAS.497.4162V} {497, 4162}

\bibitem[\protect\citeauthoryear{{Virtanen} et~al.,}{{Virtanen}
  et~al.}{2020}]{Virtanen2020}
{Virtanen} P.,  et~al., 2020, \mn@doi [Nature Methods]
  {10.1038/s41592-019-0686-2}, \href
  {https://ui.adsabs.harvard.edu/abs/2020NaMet..17..261V} {17, 261}

\bibitem[\protect\citeauthoryear{{Wenger} et~al.,}{{Wenger}
  et~al.}{2000}]{Wenger2000}
{Wenger} M.,  et~al., 2000, \mn@doi [\aaps] {10.1051/aas:2000332}, \href
  {https://ui.adsabs.harvard.edu/abs/2000A&AS..143....9W} {143, 9}

\bibitem[\protect\citeauthoryear{{Wetzel}, {Hayward}, {Sanderson}
  et~al.}{{Wetzel} et~al.}{2023}]{Wetzel2023}
{Wetzel} A.,  {Hayward} C.~C.,  {Sanderson} R.~E.,   et~al., 2023, \mn@doi
  [\apjs] {10.3847/1538-4365/acb99a}, \href
  {https://ui.adsabs.harvard.edu/abs/2023ApJS..265...44W} {265, 44}

\bibitem[\protect\citeauthoryear{{Wetzel} et~al.,}{{Wetzel}
  et~al.}{2025}]{Wetzel2025}
{Wetzel} A.,  et~al., 2025, \mn@doi [arXiv e-prints]
  {10.48550/arXiv.2508.06608}, \href
  {https://ui.adsabs.harvard.edu/abs/2025arXiv250806608W} {p. arXiv:2508.06608}

\bibitem[\protect\citeauthoryear{{Wheeler} et~al.,}{{Wheeler}
  et~al.}{2017}]{Wheeler2017}
{Wheeler} C.,  et~al., 2017, \mn@doi [\mnras] {10.1093/mnras/stw2583}, \href
  {https://scixplorer.org/abs/2017MNRAS.465.2420W} {465, 2420}

\bibitem[\protect\citeauthoryear{{Wilson} et~al.,}{{Wilson}
  et~al.}{2019}]{Wilson2019}
{Wilson} J.~C.,  et~al., 2019, \mn@doi [\pasp] {10.1088/1538-3873/ab0075},
  \href {https://ui.adsabs.harvard.edu/abs/2019PASP..131e5001W} {131, 055001}

\bibitem[\protect\citeauthoryear{{Xiang} \& {Rix}}{{Xiang} \&
  {Rix}}{2022}]{Xiang2022}
{Xiang} M.,  {Rix} H.-W.,  2022, \mn@doi [\nat] {10.1038/s41586-022-04496-5},
  \href {https://scixplorer.org/abs/2022Natur.603..599X} {603, 599}

\bibitem[\protect\citeauthoryear{{Zeng}, {Wang}, {Gao}  \& {Yang}}{{Zeng}
  et~al.}{2024}]{Zeng2024}
{Zeng} G.,  {Wang} L.,  {Gao} L.,   {Yang} H.,  2024, \mn@doi [\mnras]
  {10.1093/mnras/stae1651}, \href
  {https://scixplorer.org/abs/2024MNRAS.532.2558Z} {532, 2558}

\bibitem[\protect\citeauthoryear{{Zonca}, {Singer}, {Lenz}, {Reinecke},
  {Rosset}, {Hivon}  \& {Gorski}}{{Zonca} et~al.}{2019}]{Zonca2019}
{Zonca} A.,  {Singer} L.,  {Lenz} D.,  {Reinecke} M.,  {Rosset} C.,  {Hivon}
  E.,   {Gorski} K.,  2019, \mn@doi [The Journal of Open Source Software]
  {10.21105/joss.01298}, \href
  {https://ui.adsabs.harvard.edu/abs/2019JOSS....4.1298Z} {4, 1298}

\bibitem[\protect\citeauthoryear{{de Graaff} et~al.,}{{de Graaff}
  et~al.}{2024}]{deGraaff2024}
{de Graaff} A.,  et~al., 2024, \mn@doi [\aap] {10.1051/0004-6361/202347755},
  \href {https://scixplorer.org/abs/2024A&A...684A..87D} {684, A87}

\bibitem[\protect\citeauthoryear{{de los Reyes}, {Kirby}, {Zhuang}, {Steidel},
  {Chen}  \& {Wheeler}}{{de los Reyes} et~al.}{2023}]{delosReyes2023}
{de los Reyes} M. A.~C.,  {Kirby} E.~N.,  {Zhuang} Z.,  {Steidel} C.~C.,
  {Chen} Y.,   {Wheeler} C.,  2023, \mn@doi [\apj] {10.3847/1538-4357/acd189},
  \href {https://scixplorer.org/abs/2023ApJ...951...52D} {951, 52}

\bibitem[\protect\citeauthoryear{{van der Marel} \& {Kallivayalil}}{{van der
  Marel} \& {Kallivayalil}}{2014}]{vanderMarel2014}
{van der Marel} R.~P.,  {Kallivayalil} N.,  2014, \mn@doi [\apj]
  {10.1088/0004-637X/781/2/121}, \href
  {https://ui.adsabs.harvard.edu/abs/2014ApJ...781..121V} {781, 121}

\bibitem[\protect\citeauthoryear{{van der Marel}, {Alves}, {Hardy}  \&
  {Suntzeff}}{{van der Marel} et~al.}{2002}]{vanderMarel2002}
{van der Marel} R.~P.,  {Alves} D.~R.,  {Hardy} E.,   {Suntzeff} N.~B.,  2002,
  \mn@doi [\aj] {10.1086/343775}, \href
  {https://ui.adsabs.harvard.edu/abs/2002AJ....124.2639V} {124, 2639}

\makeatother
\end{thebibliography}

\bsp
\label{lastpage}
\end{document}